\documentclass[acmsmall]{acmart}

\acmJournal{PACMPL}
\acmVolume{1}
\acmNumber{submission to POPL} 
\acmArticle{1}
\acmYear{2024}
\acmMonth{1}
\acmDOI{} 
\startPage{1}

\setcopyright{none}

\usepackage{booktabs}   
\usepackage{subcaption} 
\usepackage{listings}
\usepackage{lscape}

\usepackage{mathpartir}
\usepackage{amsmath}
\usepackage{amsthm}
\newtheorem{theorem}{Theorem}
\newtheorem{corollary}[theorem]{Corollary}

\newcommand{\figref}[1]{Fig.~\ref{#1}}

\newcommand{\denote}[1]{\llbracket #1 \rrbracket}
\newcommand{\With}{\mathrm{With} }
\newcommand{\Require}{\mathrm{Require}}
\newcommand{\Ensure}{\mathrm{Ensure}}
\newcommand{\Assert}{\mathrm{Assert} \ }

\newcommand{\entails}{\vDash}

\newcommand{\hoare}[6]{\{#1\}\,#2\,\left\{#3, [#4, #5, #6]\right\}}
\newcommand{\hoareabbr}[4]{\left\{#1\right\}\,#2\,\left\{#3, [\vec{#4}]\right\}}
\newcommand{\hoaretriple}[3]{\left\{#1\right\}\,#2\,\left\{#3\right\}}
\newcommand{\EX}[3]{\exists #1:#2.\ #3}
\newcommand{\cskip}{\mathsf{skip}}
\newcommand{\cifthenelse}[3]{\mathsf{if}\ (#1)\ #2\ \mathsf{else}\ #3}

\newcommand{\cloop}[2]{\mathsf{loop}\ (#2)\ #1}
\newcommand{\cbreak}{\mathsf{break}}
\newcommand{\ccontinue}{\mathsf{continue}}
\newcommand{\creturn}{\mathsf{return}}
\newcommand{\cvar}[1]{\mathsf{#1}}

\newcommand{\coninv}[1]{#1_{\mathrm{con}}}
\newcommand{\brkcnd}[1]{#1_{\mathrm{brk}}}
\newcommand{\concnd}[1]{#1_{\mathrm{con}}}
\newcommand{\retcnd}[1]{#1_{\mathrm{ret}}}
\newcommand{\cincr}[1]{#1_{\mathrm{incr}}}
\newcommand{\pure}{\mathsf{pure}}
\newcommand{\se}{\mathsf{ae}}

\newcommand{\hasvalue}{\mathbin{\Downarrow}}

\renewcommand{\implies}{\Rightarrow}
\newcommand{\rulename}[1]{\textsc{#1}}
\renewcommand{\ll}{\mathrm{ll}}
\newcommand{\rev}{\mathrm{rev}}

\newcommand{\cassign}[2]{{#1} := {#2}}
\newcommand{\ccall}[3]{{#1} := {#2}({#3})}
\newcommand{\cassert}[1]{\textsf{assert} \ {#1}}
\newcommand{\ExGivensymbol}{\mathrm{ExGiven}}
\newcommand{\ExGiven}[4]{\ExGivensymbol\ #1:#2,\ \{{#3}\} \ {#4}}

\newcommand{\wand}{\mathrel{-\mkern-6mu\ast}}

\newcommand{\sepcon}{\ast}
\newcommand{\TT}{\textsf{True}}

\newcommand{\funspecabbr}[3]{\With\ [#1].\ \Require \ \{#2\} \ \Ensure\ \left\{#3\right\}}

\newcommand{\resat}[2]{{#1} @ {#2}}

\newcommand{\EXabbr}[2]{\exists #1.\ #2}

\newcommand{\join}{\oplus}
\newcommand{\bupd}{|\mkern-4mu\Rightarrow}
\newcommand{\bupdd}{|\mkern-8mu\Rightarrow}
\newcommand{\hoareabbrr}[4]{\{#1\}\,#2\,_{\bupd}\left\{#3, [\vec{#4}]\right\}}

\newcommand{\semaxpre}{\mathrm{HeadHypo}}
\newcommand{\semaxpost}{\mathrm{TailHypo}}
\newcommand{\semaxpath}{\mathrm{FullHypo}}
\newcommand{\semaxatom}{\mathrm{AssnFreeHypo}}
\newcommand{\semaxsplit}{\mathrm{AllHypo}}

\newcommand{\semaxprea}{\mathrm{HeadHypo}}
\newcommand{\semaxposta}{\mathrm{TailHypo}}
\newcommand{\semaxpatha}{\mathrm{FullHypo}}
\newcommand{\semaxatoma}{\mathrm{AssnFreeHypo}}

\newcommand{\eraseannot}[1]{#1\mkern-4mu\Downarrow}
\newcommand{\splitfun}[1]{\mathsf{split}#1}
\newcommand{\nil}{{\emptyset}}
\newcommand{\singleton}[1]{{\{#1\}}}
\newcommand{\app}{ \mkern-4mu+\mkern-8mu+\mkern+4mu }
\newcommand{\append}{ \cup }
\newcommand{\conn}{ \cdot }

\usepackage{mathpartir}

\usepackage{tikz-cd}
\usepackage[framemethod=tikz]{mdframed}
\usetikzlibrary{positioning, shapes.geometric, fit, calc}

\usepackage{color}
\usepackage{amsmath}

\usepackage{xspace}

\newif\ifhasappendix
\hasappendixtrue 
\newcommand{\seeappendix}{\ifhasappendix{the appendix}\else{the extended version of the paper}\fi \xspace}

\begin{document}

\title[VST-A]{VST-A: A Foundationally Sound Annotation Verifier} 


\author{Litao Zhou}
\affiliation{
  \institution{Shanghai Jiao Tong University}           
  \country{China}
}
\affiliation{
  \institution{The University of Hong Kong}
  \country{China}
}
\email{tonyzhou0608@gmail.com}         

\author{Jianxing Qin}
\affiliation{
    \institution{Shanghai Jiao Tong University}
    \country{China}
}
\email{qdelta@sjtu.edu.cn}

\author{Qinshi Wang}
\affiliation{
    \institution{Princeton University}
    \country{United States}
}
\email{qinshiw@cs.princeton.edu}

\author{Andrew W. Appel}
\affiliation{
    \institution{Princeton University}
    \country{United States}
}
\email{appel@princeton.edu}

\author{Qinxiang Cao}
\authornote{Corresponding author.}
\affiliation{
  \institution{Shanghai Jiao Tong University}           
  \country{China}
}
\email{caoqinxiang@gmail.com}         

\begin{abstract}
Program verifiers for imperative languages such as C may be \emph{annotation-based}, in which assertions and invariants are put into source files and then checked, or tactic-based, where proof scripts separate from programs are \emph{interactively} developed in a proof assistant such as Coq.  Annotation verifiers have been more automated and convenient, but some interactive verifiers have richer assertion languages and formal proofs of soundness.   We present VST-A, an annotation verifier that uses the rich assertion language of VST, leverages the formal soundness proof of VST, but allows users to describe functional correctness proofs intuitively by inserting assertions.

VST-A analyzes control flow graphs, decomposes every C function into control flow paths between assertions, and reduces program verification problems into corresponding \emph{straightline Hoare triples}. Compared to existing foundational program verification tools like VST and Iris, in VST-A, such decompositions and reductions are allowed to be nonstructural, which makes VST-A more flexible to use.

VST-A's decomposition and reduction is defined in Coq, proved sound in Coq, and computed in a call-by-value way in Coq. The soundness proof for reduction is totally logical, independent of the complicated semantic model (and soundness proof) of
VST's Hoare triple. Because of the rich assertion language, not all reduced proof goals can be automatically checked, but the system allows users to prove residual proof goals using the full power of the Coq proof assistant.
\end{abstract}



\maketitle

\section{Introduction}
\label{sec:intro}


In the past 15 years, researchers have built several tools for program verification.
These tools are used in different ways and have their own advantages.

Interactive program verification tools, such as 
Iris~\cite{IrisProofMode, jung2018iris} and
VST~\cite{VST, VST-Floyd, VSU}, are built in 
interactive theorem provers like Coq or HOL4.
Users write
formal program correctness proofs
in the same theorem prover,
by using the lemmas and tactics of
the program verification tool.

Some of these interactive tools themselves are 
\emph{foundationally sound} (i.e., have a formal proof w.r.t. the language's
operational semantics in the proof assistant). This is especially meaningful 
for verifying real-world programs and higher-level
properties such as functional correctness. First, real-world programming 
languages are complicated. For example, it is very subtle
to determine what C programs may 
cause undefined behavior. Second, an advanced program logic for 
higher-order properties usually has a nontrivial soundness proof.
For example, VST and Iris use step-indexed semantics to interpret
impredicative assertion languages whose soundness proof is complicated.

Interactive program verification tools can also benefit from the rich logic language and the
rich proof language of theorem provers (like Coq and HOL).
These tools can easily shallow-embed higher-order
functions and predicates in their assertion languages. They also make it convenient (in specifying programs
and program assertions) to introduce additional logical connectives.
Additionally, the tactic proof languages in proof assistants are very powerful when users need to describe 
proof strategies such as proof-by-induction and proof-by-contradiction.

A different strain of program verification tools
requires programmers to write
annotations in the source code. With sufficient
annotations, tools like Dafny~\cite{dafny}, Hip/Sleek~\cite{hipsleek},
VeriFast~\cite{verifast}, Viper~\cite{viper}, Frama-C~\cite{Frama-C} and CN~\cite{Pulte23POPL} can
verify program correctness automatically. By restricting the assertion languages,
tools like CBMC~\cite{cbmc}, F-Soft~\cite{fsoft, fsoft-journal}, and Infer~\cite{infer}
can reduce the annotation overhead for programmers while preserving automation.

Compared with writing tactical proofs in a theorem prover, 
\emph{writing annotations is a much more straightforward way of 
      demonstrating that a program is correct.}
Even proofs in theory papers and proofs written completely in an interactive prover 
      are often presented in research papers as annotated programs, e.g. Reynolds's 
      first paper about separation logic uses annotations to describe separation 
      logic proofs ~\cite[page 10]{reynolds02}, and recent verification papers 
      like \citet{DBLP:journals/pacmpl/JungLPRTDJ20}'s  work of 
      extending Iris to support prophercy variables also uses annotations to 
      describe their proofs~\cite[page 7]{DBLP:journals/pacmpl/JungLPRTDJ20}.
\figref{fig:ll-reverse} shows an implementation of 
    an in-place linked list reversal and its functional correctness proof.%
\footnote{This is a separation logic \cite{reynolds02} proof.
    In other words, we use an assertion of form ``$P * Q$'' to say that 
    the memory can be split into two disjoint pieces of which one 
    satisfies $P$ and the other satisfies $Q$. $\ll(\cvar{p}, l)$ is a
    separation logic predicate that asserts on the location referenced
    by variable $p$ stores a linked list of $l$.}
The annotations on lines 3-5 describe the specification this function should satisfy: for any list of integers $l$ (the $\mathrm{With}$ clause on line \ref{code:reverse-with}), if $l$ is stored in a linked list and this link list's head pointer is passed to the function by the program variable $\cvar{p}$ (the $\mathrm{Require}$ clause on line \ref{code:reverse-require}), then the function  reverses the linked list and 
returns the new head pointer (the $\mathrm{Ensure}$ clause on line \ref{code:reverse-ensure}). The assertion on line \ref{code:reverse-assert} describes the main idea of a functional correctness proof.
It states the criteria that the program state should satisfy every time the program enters the loop body.
Assertion-annotated programs can present proofs succinctly; 
by contrast, in interactive verifiers,
key insights into program correctness easily get mixed with structural 
tactics and become lengthy proof scripts.

\begin{figure}[h]
     \begin{lstlisting}[language=C, mathescape=true, xleftmargin=2em,xrightmargin=1em, numbers=left, frame = {TB}, escapechar=|,]
struct list {unsigned head; struct list *tail;};
struct list *reverse (struct list *p) {
     /*@ $\With \ \ l,$ |\label{code:reverse-with}|
         $\Require \ \ \ll\left(\cvar{p}, l\right)$  |\label{code:reverse-require}|
         $\Ensure \ \ \ll\left(\cvar{ret}, \rev(l)\right)$ */  |\label{code:reverse-ensure}|
     struct list *w, *t, *v;
     w = NULL; v = p;
     while (v) {
          /*@ $\Assert \exists \ l_1 \ u \ x \ l_2. \ 
          l = \rev(l_1) \ x \ l_2 \land \ 
          \cvar{v} \mapsto (x, u) \ * \
          \ll\left(\cvar{w}, l_1\right) \ * \
          \ll\left(u, l_2\right)$ */  |\label{code:reverse-assert}|
          t = v->tail; v->tail = w; w = v; v = t; }
     return w; }
     \end{lstlisting}
     \caption{Annotations for verifying linked-list reversal}
     \label{fig:ll-reverse}
\end{figure}

In this paper, we demonstrate how to combine the benefits of interactive tools and 
annotation verifiers.
We present VST-A, a foundationally sound verification tool, 
that is implemented on top of VST.
VST-A enjoys rich assertion languages and flexible proof strategies,
and it allows users to write readable
assertion annotations directly as comments 
exactly as in \figref{fig:ll-reverse}.

\begin{figure}[ht]
     \begin{center}
        {\footnotesize
               \includegraphics[scale=0.6]{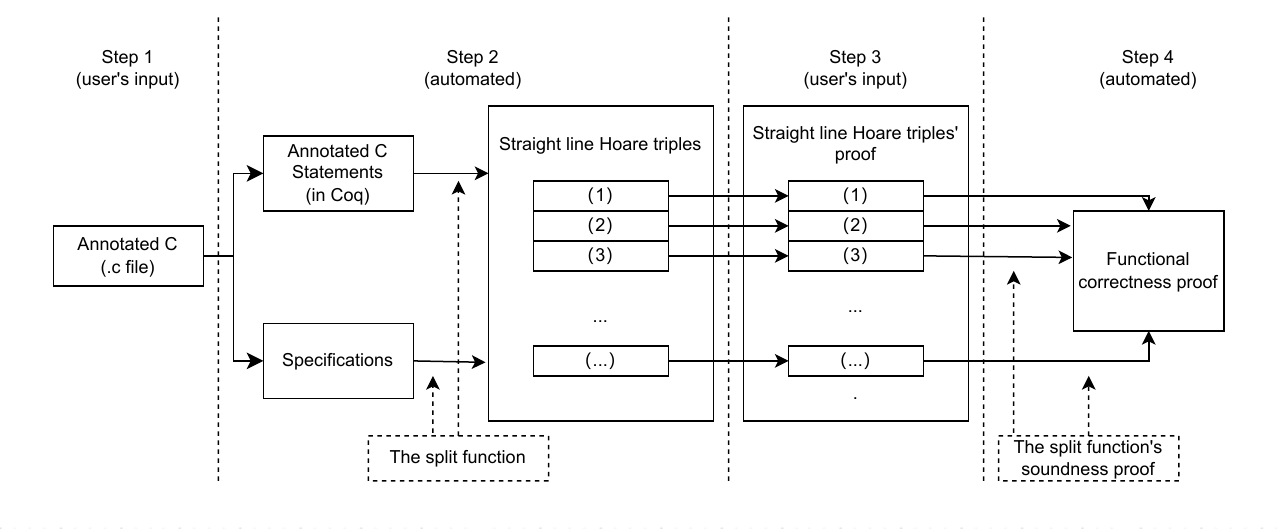}
        }
     \caption{Verification workflow in VST-A}
     \label{fig:workflow}
     \end{center}
   \end{figure}

We illustrate the VST-A workflow in 
\figref{fig:workflow}:
(1) Users first provide a C program with assertion annotations.
(2a) Our front-end parser will then convert the source code into
ClightA, the Coq representation of this annotated C language.
(2b) Next, the C program's functional correctness is reduced to smaller proof 
    goals using the annotations. Specifically,
a \emph{split function} accepts a ClightA program and its
pre-/post-conditions as input, and returns a set of straightline 
Hoare triples, each of which consists of a sequence of
primary statements\footnote{
     We refer to single assignment and function call statements
     as primary statements in this paper.
}
and/or \textbf{\textsf{assume}} commands.
For example, \figref{fig:ll-reverse-split} shows the split result 
of the \textsf{reverse} function in \figref{fig:ll-reverse}: 
four triples are returned as verification goals.
As illustrated by the control flow graph in \figref{fig:ll-reverse-cfg}, 
the functionality of this \emph{split function} is natural; 
it computes all of the control flow paths that are 
separated by assertion annotations 
in the source program.
(3) Finally, users are left to prove each straightline 
Hoare triple in the split result. (4) The VST-A soundness theorem
ensures the correctness of the original program if all of the paths have been verified.

\begin{figure}[h]
     \begin{minipage}[b]{0.5\textwidth}
          \small{
               \input{figures/split_example.tex}
          }
          \caption{Split results for verifying linked-list reversal}
          \label{fig:ll-reverse-split}
     \end{minipage}
     \begin{minipage}[b]{0.45\textwidth}
          \centering
          \small{
               \begin{tikzpicture}[node distance=20pt]

    \node[align=center]
        (require) { \{ Require \} };

    \node[draw, below=of require,  align=left, minimum width=0.3\textwidth]
        (code1) { \footnotesize
\begin{lstlisting}[language=C, mathescape=true, xleftmargin=0em,xrightmargin=0em]
struct list *w, *t, *v;
w = NULL; v = p;
\end{lstlisting}
        };

    \node[draw, below=of code1, align=left, diamond, shape aspect=1.5, minimum width=0.2\textwidth]
        (code_cond) {\footnotesize
\begin{lstlisting}[language=C, mathescape=true, xleftmargin=0em,xrightmargin=0em]
v == NULL?
\end{lstlisting}
        };

    \node[below=of code_cond, align=center]
        (assert) { \{ Assert \} };

    \node[draw, below=of assert, align=left, minimum width=0.3\textwidth]
        (code2) {\footnotesize
\begin{lstlisting}[language=C, mathescape=true, xleftmargin=0em,xrightmargin=0em]
t = v->tail;
v->tail = w;
w = v; v = t;
\end{lstlisting}
        };

    \node[draw, below=40pt of code2, align=left, minimum width=0.3\textwidth]
        (code3) { \footnotesize
\begin{lstlisting}[language=C, mathescape=true, xleftmargin=0em,xrightmargin=0em]
return w;
\end{lstlisting}
        };

    \node[below=of code3, align=center]
        (ensure) { \{ Ensure \} };

    \coordinate [right=of code_cond] (code_cond_coord);
    \coordinate [right=of code3] (code3_coord);
    \coordinate [left=of code_cond] (code_cond_coord2);
    \coordinate [below=18pt of code2] (code2_coord);

    \draw[->, thick] (require) -- (code1);
    \draw[-, orange] ([xshift=2ex] require.south) -- 
        ([xshift=2ex] code1.north);
    \draw[-, red] ([xshift=-2ex] require.south) -- 
        node [left] { (1) } 
        ([xshift=-2ex] code1.north);

    \draw[->, thick] (code1) -- (code_cond);
    \draw[-, orange] ([xshift=2ex] code1.south) -- ([xshift=2ex] code_cond.north);
    \draw[-, red] ([xshift=-2ex] code1.south) -- ([xshift=-2ex] code_cond.north);

    \draw[->, thick] (code_cond) -- (assert);
    \draw[->, brown] ([xshift=2ex] code_cond.south) -- ([xshift=2ex] assert.north);
    \draw[->, red] ([xshift=-2ex] code_cond.south) -- ([xshift=-2ex] assert.north);

    \draw[->, thick] (assert) -- (code2);
    \draw[-, brown] ([xshift=2ex] assert.south) -- ([xshift=2ex] code2.north);
    \draw[-, blue] ([xshift=-2ex] assert.south) -- ([xshift=-2ex] code2.north);

    \draw[->, thick] (code2) --
        (code2_coord) --
        (code2_coord -| code_cond_coord2) --
        (code_cond_coord2) --
        (code_cond);
    \draw[-, brown] ([xshift=2ex] code2.south) --
        ([xshift=2ex, yshift=-2ex]code2_coord) --
        ([xshift=-2ex, yshift=-2ex]code2_coord -| code_cond_coord2) --
        node [left] { (3) } 
        ([xshift=-2ex, yshift=2ex]code_cond_coord2) --
        ([yshift=2ex] code_cond.west);
    \draw[-, blue] ([xshift=-2ex] code2.south) --
        ([xshift=-2ex, yshift=2ex]code2_coord) --
        ([xshift=2ex, yshift=2ex]code2_coord -| code_cond_coord2) --
        ([xshift=2ex, yshift=-2ex]code_cond_coord2) --
        ([yshift=-2ex] code_cond.west);

    \draw[->, thick] (code_cond) --
        (code_cond_coord) --
        (code_cond_coord|-code3_coord) --
        (code3);
    \draw[-, orange] ([yshift=2ex] code_cond.east) --
        node [above] { (2) } 
        ([xshift=2ex, yshift=2ex]code_cond_coord) --
        ([xshift=2ex, yshift=-2ex]code_cond_coord|-code3_coord) --
        ([yshift=-2ex]code3.east);
    \draw[-, blue] ([yshift=-2ex] code_cond.east) --
        ([xshift=-2ex, yshift=-2ex]code_cond_coord) --
        node [left] { (4) } 
        ([xshift=-2ex, yshift=2ex]code_cond_coord|-code3_coord) --
        ([yshift=2ex]code3.east);

    \draw[->, thick] (code3) -- (ensure);
    \draw[->, orange] ([xshift=2ex] code3.south) -- ([xshift=2ex] ensure.north);
    \draw[->, blue] ([xshift=-2ex] code3.south) -- ([xshift=-2ex] ensure.north);

\end{tikzpicture}
          }
          \caption{Control flow graph of linked-list reversal}
          \label{fig:ll-reverse-cfg}
     \end{minipage}
\end{figure}

In summary, the contributions of this paper are:
\begin{enumerate}
     \item \textbf{A new framework for program verification} that combines the 
     benefits of interactive provers and the readability of
     annotated programs.
     \item \textbf{A formal language for annotated programs}: We define ClightA,
     a formal language for annotated C programs, as a method of describing how
     the functional correctness proof for a large program can be 
     reduced.
     \item \textbf{A control-flow-based verification splitting algorithm}:
     The VST-A proof reduction framework uses a split function
     that is implemented in Coq and proved sound w.r.t. the VST program logic.
     We believe this split algorithm and its soundness proof are general 
     and can be applied to other Hoare-style imperative
     verification tools as well.
\end{enumerate}

In the rest part of this paper, we will introduce background information about VST-A in \S\ref{sec:backgrnd}.
We will present our annotation-based proof language and analyze its expressiveness in \S\ref{sec:frontend}.
We will define the split function and prove it sound in \S\ref{sec:split_and_sound}.
We will discuss the connection between our soundness proof and Hoare logic's conjunction rule in \S\ref{sec:conj}.
We put statistics of VST-A verification examples in \S\ref{sec:eval}.
In the end, we will discuss related works in \S\ref{sec:relate} and conclude in \S\ref{sec:concl}.

\section{Background}
\label{sec:backgrnd}

We used Coq~\cite{COQ} and VST to implement our annotation verifier, VST-A.
VST is an interactive program verification tool that is built in Coq.
Its primary components are
\begin{enumerate}
    \item Verifiable C (\S\ref{sec:hoarelogic}, \S\ref{sec:wp}), an impredicative higher-order concurrent separation logic that is
    defined for an abstract C language called Clight (\S\ref{sec:clight}),
    \item VST-Floyd (\S\ref{sec:forward}) ~\cite{VST-Floyd}, a proof automation system for forward symbolic execution-based
    verification that efficiently applies VST to real-world C program verification,
    \item A machine-checked soundness proof of Verifiable C in terms of CompCert Clight semantics. 
    Together with the correctness proof of the verified C compiler\textemdash CompCert~\cite{Leroy2009}, 
    we can obtain the foundational soundness of VST-A w.r.t. the assembly language.
\end{enumerate}

\subsection{Clight: abstract C language}
\label{sec:clight}

We reason about C programs using CompCert Clight's syntax and semantics. 
\figref{fig:ClightAST} shows a simplified version of its syntax.%
\footnote{
    VST-A does not support goto statements, since they are not supported by VST's program logic.
    VST-A does not support switch statements for now, but it will be easy to add that to VST-A in the future.
    VST-A does not yet support CompCert's special calls to built-in functions;
    these are rarely used in the source language.}
Clight expressions are side-effect free.
CompCert Clight distinguishes assignment statements and function call statements
    from other statements, and we refer to them as \emph{primary statements}, since
    they are the basic building blocks for our VST-A development.

CompCert Clight uses $\cloop{c}{\cincr{c}}$ as a general way to describe loops,
    and it is equivalent to $\mathsf{for}\ (;;\cincr{c})\ \{c\}$.
Three kinds of loops in C language, namely for, while, and do-while loops,
    can be expressed using this general loop statement
    (along with break and continue statements).
In the paper presentation, we assume that return statements do not return a value,
while our Coq development does handle return statements with return values.%
\footnote{
    VST handles return values by a reserved variable called $\cvar{ret}$, as we illustrated
    in \figref{fig:ll-reverse-split}. Supporting return statements with return values does not 
    cause significant difficulties in our development.
}

\begin{figure}[t]
$$\begin{array}{rrl}
    \text{expression}:      &e := & \cdots \\
    \text{primary statement}:&c_p := & \cassign{e_1}{e_2} \ \mid \ \ccall{e}{f}{\vec{e}} \\
    \text{Clight statement}: &c :=& c_p  \ \mid \  c_1; c_2 \ \mid \ \cifthenelse{e}{c_1}{c_2} \ \mid \ \cloop{c_1}{c_2} \\
                            &\mid&  \cskip \ \mid \ \cbreak \ \mid \ \ccontinue \ \mid \ \creturn 
\end{array}$$
    \caption{Clight: abstract C language}
    \label{fig:ClightAST}
\end{figure}

\begin{figure*}[t]
\centering
\begin{mathpar}
    \inferrule*[left=Semax-Seq]{
        \hoareabbr{P}{c_1}{R}{Q'} \and
        \hoareabbr{R}{c_2}{Q}{Q'}
    } {
        \hoareabbr{P}{c_1;c_2}{Q}{Q'}
    }
    \and
    \inferrule*[left=Semax-Loop]{
        \hoare{I}{c}{\coninv{I}}{Q}{\coninv{I}}{\retcnd{Q}} \and
        \hoare{\coninv{I}}{\cincr{c}}{I}{Q}{\bot}{\retcnd{Q}}
    } {
        \hoare{I}{\cloop{c}{\cincr{c}}}{Q}{\brkcnd{Q}}{\concnd{Q}}{\retcnd{Q}}
    }
\end{mathpar}
\caption{Representative proof rules of C Hoare logic (Part I: compositional rules)}
\label{fig:choarelogic}
\end{figure*}

\subsection{Hoare logic for C programs}
\label{sec:hoarelogic}

VST-A reuses VST's Hoare logic rules, which are known as Verifiable C.
VST's Hoare judgment extends the postcondition
into four parts to address control flow instructions such as $\cbreak$,
$\ccontinue$ and $\creturn$. A judgment
$$\hoare{P}{c}{Q}{\brkcnd{Q}}{\concnd{Q}}{\retcnd{Q}}$$ can be interpreted as
starting from a program state that satisfies $P$. After executing $c$,
if the statement exits normally, the program state satisfies $Q$. Similarly,
the program state should satisfy $\brkcnd{Q}$, $\concnd{Q}$, and $\retcnd{Q}$
when the statement exits with $\cbreak$, $\ccontinue$, or $\creturn$,
respectively. We use $\left[\vec{Q}\right]$ as an abbreviation of the last three postconditions.

In Verifiable C, most of the compositional rules are standard. \figref{fig:choarelogic} shows some representative ones.%
\footnote{A full list of rules can be found in \seeappendix.}
To fit the general loop syntax of Clight,
the \rulename{Semax-Loop}
    rule has two invariants, loop invariant $I$ and continue invariant $\coninv{I}$.
    The loop invariant $I$ is required to hold before each iteration,
    and $\coninv{I}$ is required to hold at $\ccontinue$
    or before $\cincr{c}$ statements.
Verifiable C's proof rules for primary statements are less important --- VST 
    provides verified forward symbolic execution (\S\ref{sec:forward}), and 
    thus VST's users do not need to use those rules directly.
All proof rules in Verifiable C
    are proved sound foundationally w.r.t. the CompCert Clight semantics,
    and symbolic execution applies the proof rules.
VST also provides some useful derived rules,
    a few of which are shown in \figref{fig:choarelogic_derived}. 
The \rulename{Seq-Assoc} rule reassociates sequential compositions, and
rules \rulename{Extract-Exists} and \rulename{Extract-Pure}
    introduce variables and propositions from precondition to context, respectively.

\begin{figure*}[h]
\centering
\begin{mathpar}
    \inferrule*[left=Extract-Pure]{
        \pure(P_\pure) \and
        P_\pure \Rightarrow \hoareabbr{P}{c}{Q}{Q'}
        } {
        \hoareabbr{P_\pure\land P}{c}{Q}{Q'}
        }
        \and
    \inferrule*[left=Extract-Exists]{
        \forall \ (x:A). \ \hoareabbr{P}{c}{Q}{Q'}
    } {
        \hoareabbr{\EX{x}{A}{P}}{c}{Q}{Q'}
    }
    \and
    \inferrule*[left=Seq-Assoc]{
        \hoareabbr{P}{c_1;(c_2;c_3)}{Q}{Q'}
    } {
        \hoareabbr{P}{(c_1;c_2);c_3}{Q}{Q'}
    }
\end{mathpar}
\caption{Derived rules from C Hoare logic}
\label{fig:choarelogic_derived}
\end{figure*}

\subsection{Inversion rules and weakest preconditions}
\label{sec:wp}

The VST program logic is higher-order, i.e., assertions can be quantified over
assertions, so that one can state the following inversion lemmas,
which have already been proven in VST.%
\footnote{Readers who knew that
VST's separation Hoare logic was proved sound with a semantic proof
in a shallow-embedded style may be surprised that it is possible to prove
inversion lemmas.  But in fact, several years ago the VST-Floyd 
developers layered
a deep-embedded Hoare logic over the shallow-embedded Hoare logic
just so that useful lemmas of this kind can be supported.}

\begin{lemma}[Inversion on sequencing]
    \label{lemma:semax_seq_inv}
    If $\hoareabbr{P}{c_1;c_2}{Q}{Q'}$, then
    $$\hoareabbr{P}{c_1}{ \EX{R}{\textsf{assert}}{R \land \left(\hoareabbr{R}{c_2}{Q}{Q'}\right)} }{Q'}$$
\end{lemma}

\begin{lemma}[Inversion on if-branching]
    \label{lemma:inversion_if}
    If $\hoareabbr{P}{\cifthenelse{b}{c_1}{c_2}}{Q}{Q'}$, then
    $$
    \begin{array}{l}
       {P} \entails \exists P': \textsf{assert}, \ P'  \land \left(\hoareabbr{P'\land b \neq 0 }{c_1}{Q}{Q'}\right) \land \left(\hoareabbr{P'\land b = 0 }{c_2}{Q}{Q'}\right) \\
    \end{array}
    $$    
\end{lemma}

It is worth mentioning that the normal postcondition appearing in the inversion on sequencing is VST's representation of weakest precondition, i.e.
$$ \text{wp}\left(c, Q, \left[\vec{Q'}\right]\right) \triangleq \EX{R}{\textsf{assert}}{R \land \left(\hoareabbr{R}{c}{Q}{Q'}\right)}.$$
This definition of weakest precondition satisfies basic properties like the following:
\begin{theorem}
    \label{thm:wp}
    $P \entails \text{wp}\left(c, Q, \left[\vec{Q'}\right]\right)$ if and only if $\hoareabbr{P}{c}{Q}{Q'}$.
\end{theorem}
Thus, lemma \ref{lemma:semax_seq_inv} can be restated as:
$$\hoareabbr{P}{c_1;c_2}{Q}{Q'} \ \text{iff.} \ \hoareabbr{P}{c_1}{\text{wp}\left(c_2, Q, \left[\vec{Q'}\right]\right)}{Q'}.$$
We utilize higher-order assertions in our VST-A development.

\subsection{Forward symbolic execution}

\label{sec:forward}
VST’s forward verification tactics enable users to obtain the strongest postcondition of a sequence of primary statements automatically.
For example, in the verification of path (3) in \figref{fig:ll-reverse-split},
the symbolic assignment executor $\se(P, c)$ for precondition $P$ and 
statement $c$ can compute the following (where ll is the linked-list predicate):

$$
\begin{array}{rl}
    P :=& l = \rev(l_1) \ x \ l_2 \ \land \
    \cvar{v} \mapsto (x, u) \ * \ 
  \ll\left(\cvar{w}, l_1\right) \ * \
    \ll\left(u, l_2\right) \\
    c :=& \textsf{t = v->tail} \\
    \se(P, c) =& l = \rev(l_1) \ x \ l_2 \land \ 
        \cvar{t} = u \ \land \
        \cvar{v} \mapsto (x, u) \ * \
      \ll\left(\cvar{w}, l_1\right) \ * \ 
        \ll\left(u, l_2\right)
\end{array}
$$

The assignment executor may fail,
    if the precondition $P$ cannot guarantee that $c$ will run safely
    or $P$ is not in a good form\footnote{
        Generally speaking, to ensure that an assignment executor $\se(P, c)$ always succeeds,
        $P$ should be in the form of a symbolic heap assertion.
        When $c$ is a load/store statement, there should be an explicit mapsto predicate
        for the manipulated variables in the separating conjunction clauses.
        Otherwise, one  might need to apply the rule of consequence
        (and prove an entailment) to put $P$ into that form.
    } such that the symbolic executor can execute $c$,
but we have found if users write
assertions in their annotations, corresponding to correct proofs,
    the symbolic executor can run through the entire straightline Hoare triple.
Users are left to prove the entailment from the inferred strongest
    postcondition (of the straight-line code)
    to the specified postcondition (arising from the ``next'' annotation,
    according to the split function).
VST-Floyd also provides useful tactics for solving the entailment problem. 
Residual proof goals may be generated if some entailments cannot be 
    automatically proven, and
    users can write their own flexible Coq proof scripts to address them.
In summary, with the help of VST-Floyd, the back-end verification of the split
    results obtained by VST-A can be largely automated.




\section{VST-A front end}
\label{sec:frontend}

\subsection{Annotated C programs and internal representations}

VST-A requires users to describe C function specifications and C programs' functional correctness proofs by writing annotations in C programs.
Specifically, a function specification in VST-A is always in a 
\textsf{/*@ $\With$ ... $\Require$ ... $\Ensure$ ... */} form and 
$$\With \ [\vec{x}:\vec{A}] \ \Require \ P(\vec{x}) \ \Ensure \ Q(\vec{x})$$ 
represents a parameterized pre-/postcondition, i.e. it states that for any list of values $\vec{x}$ of  type $\vec{A}$, if the initial program state satisfies 
$P(\vec{x})$ then the C function can be safely executed 
(no C undefined behavior will happen). If it terminates, 
the ending state satisfies $Q(\vec{x})$.

For functional correctness proofs, 
    users can describe the main proof skeleton by inserting assertions (including loop invariants) and ``given'' annotations in the source program.
We formally define this annotated C language (\figref{fig:AClightAST}), namely ClightA, and implement a front-end parser that converts annotated
    C programs into the ClightA abstract syntax.
Compared with the Clight syntax, ClightA has two new components:
    assertions and $\ExGivensymbol$ structures.

\begin{figure}[t]
$$\begin{array}{c}
    \begin{array}{rrl}
        \text{assertion}:       &P := & \cdots\\
        \text{ClightA statements}: &C :=& c_p  \ \mid \  C_1; C_2 \ \mid \ \cifthenelse{e}{C_1}{C_2} \ \mid \ \cloop{C_1}{C_2} \\
        &\mid&  \cskip \ \mid \ \cbreak \ \mid \ \ccontinue \ \mid \ \creturn    \\
                                &\mid& \colorbox{gray!10}{$\cassert{P} \ \mid \ \ExGiven{x}{A}{P(x)}{C}$} \\
        \text{Annotation erasing}: & \eraseannot{C} ~\,  \in & \text{Clight statement} 
    \end{array}
\end{array}$$
        \caption{ClightA: abstract C language for annotated programs}
        \label{fig:AClightAST}
\end{figure}

As for assertions,
    users can insert them anywhere by writing \textsf{/*@ Assert ... */}
    in the source program, which is directly converted into a leaf node in
    the ClightA syntax tree. 
Annotating loop structures 
    with invariants is \emph{not compulsory} in VST-A, as mentioned
    above, but users can still
    write loop invariants as \textsf{/*@ Inv ... */} in C source files,
    to distinguish from assertions before loops.
Our front-end automatically converts such annotations
    into the general syntax of ClightA.

As for the $\ExGivensymbol$ structures,
    users can use a combination of \textsf{/*@ Assert $\exists$ x, ... */}
    and \linebreak \textsf{/*@ Given x */} to represent logical variable introduction
    in a Hoare logic proof.
In a typical goal-directed proof strategy, one can
    extract existential variables from the precondition 
    into the proof context (see \textsc{Extract-Exist} in \figref{fig:choarelogic_derived}). 
Then assertions that appear later in the focused proof can 
    refer to the extracted variables as ordinary Coq assumptions.
VST-A supports this proof method by defining
    the ``$\ExGiven{x}{A}{P(x)}{C}$'' syntax.
The syntax indicates that assertion $P$ is existentially quantified by logical variable $x$.
Moreover, the inner ClightA statement is also quantified by $x$, so
   within the (annotated) assertions of $C$, one can mention $x$.

\figref{fig:examplefrondend} is a comparison between annotated C programs and ClightA syntax,
showing how our front-end parser translates annotations into AST constructions.

    \begin{figure}[h]
        \begin{minipage}[b]{0.45\textwidth}
             \small{
                  \input{figures/annotated_c_example.tex}
             }
        \end{minipage} \ \ \ 
        \begin{minipage}[b]{0.45\textwidth}
             \centering
             \small{
                  \input{figures/annotated_clight_example.tex}
             }
        \end{minipage}
        \caption{Annoated C program v.s. ClightA syntax}
        \label{fig:examplefrondend}
    \end{figure}

\subsection{Expressiveness: describing Hoare logic proofs in VST-A}

It is not surprising that the ClightA language is expressive enough for describing the main structures of Hoare logic proofs.
Hoare logic proof rules decompose the verification target into smaller ones (e.g., as in \figref{fig:choarelogic} in \S\ref{sec:backgrnd}).
Describing such proofs in VST-A is natural.
For example, to apply the sequence rule \textsc{Semax-Seq}, we can insert the middle condition as an assertion annotation into the C program.

\subsection{Expressiveness: describing interactive proofs in VST-A}

ClightA is also expressive enough to describe interactive proofs used by existing verification tools like VST.
In most cases, these tools try to find proof rules to apply 
using a \emph{goal-directed strategy}. For example:
\begin{itemize}
\item
    In order to verify a Hoare triple of form
    $ \{P\} \textsf{x = }e; c \left\{Q, \left[\vec{Q}'\right]\right\}$
    in VST, its forward symbolic execution tactic applies the sequence rule \textsc{Semax-Seq} and uses the strongest postcondition of $P$ and \textsf{x = e} as the middle condition.
    VST-A's users do not need to write any annotation to describe such proof strategy.
\item
    In order to verify a Hoare triple of form
    $$ \{P\} \textsf{while ($e$) \{ $c_1$ \}}; c_2 \left\{Q, \left[\vec{Q}'\right]\right\}$$
    in VST (suppose no break statements appear in $c_1$), it asks users to provide a loop invariant $I$ and then $I \land e = 0$ will be used as the middle condition between the loop and $c_2$.
    In VST-A, the proof effort is similar. Users only need to provide such a loop invariant $I$ in annotated C programs.
\item
    In order to verify a Hoare triple 
    whose precondition is existentially quantified, one will typically
    extract that existential variable into the proof context.
    The ``given'' annotation in VST-A describes this proof strategy.
\end{itemize}

Besides these goal-directed proof steps, users of interactive program-verification tools may use the rule of consequence anywhere in the middle of a proof, to replace a precondition with a weaker assertion (\figref{fig:exp-given} shows an example).
Correspondingly, users of VST-A can write assertion annotation to ``invoke'' the rule of consequence.

\begin{figure}[ht]
  \begin{center}
  $
    \begin{array}{l}
        \left\{ \ll\left(\cvar{p}, l\right) \ \land \ \cvar{p} \neq \textsf{NULL}  \right\} \\
        \textsf{p -> head = 0} \\
        \left\{ \exists l'. \ll\left(\cvar{p}, l'\right) \right\} \\
    \end{array}
  $
  can be rewritten as~~~
  $
    \begin{array}{l}
        \left\{ \exists \ q \ x \ l'.  \cvar{p} \mapsto (x, q) \sepcon \ll\left(q, l'\right) \right\} \\
        \textsf{p -> head = 0} \\
        \left\{ \exists l'. \ll\left(\cvar{p}, l'\right) \right\} \\
    \end{array}
  $
  \caption{Example of replacing the precondition with an existentially quantified assertion.  The predicate $\ll\left(\cvar{p}, l\right)$
  means that a \textbf{l}inked \textbf{l}ist starts at address $p$
  representing (in the heads of its cons cells) the sequence $l$.
  The two preconditions above are provably equivalent.}
  \label{fig:exp-given}
  \end{center}
\end{figure}

In summary, typical interactive proofs are structural,
following the syntax tree of the C statements
(with local uses of  existential variable extraction or the rule of consequence).
The ClightA language is able to describe such structural proofs.

\subsection{Expressiveness: describing nonstructural proofs in VST-A}

ClightA can also describe nonstructural proofs, those that do not follow the C syntax tree.
Allowing users to write nonstructural proofs is a considerable convenience.
For example, in order to verify a Hoare triple of form
    $$ \{P\} \textsf{if} \ (e) \ \{ \ c_1 \ \} \ \textsf{else} \ \{ c_2 \}; c_3 \left\{Q, \left[\vec{Q}'\right]\right\}$$
    in an interactive verification tool,
    users will probably be asked to provide a join condition after the if-statetment
    (a precondition for $c_3$).
    In some cases this is appropriate and convenient, but sometimes
    it is both difficult and unnecessary.
Here are some typical scenarios.
\begin{itemize}
\item
    The if-then branch $c_1$ is the break statement.
    In this case, it suffices to prove
    $$ \{P \land e \neq 0\} c_1 \left\{\bot, \left[Q_{\text{brk}}, \bot, \bot\right]\right\} \ \ \text{and} \ \
    \{P \land e = 0\} c_2; c_3 \left\{Q, \left[\vec{Q}'\right]\right\},$$
    and users may hope to symbolically execute $c_2$ so that the middle condition between $c_2$ and $c_3$ can be generated instead of manually provided.%
    \footnote{
        VST users do not need to provide a join condition in such cases since 2018. VST implemented this feature by adding a built-in program transformation to its verification tactics.
    }

\item The statement $c_3$ is very short.
    In this case, verifying $c_3$ twice can be less work than writing down a join condition.

\item Proving $c_3$ functionally correct needs very different proof strategy when $e$ has a different boolean value. In this case, a join condition does not help to reduce the workload: in effect, one is case-splitting on $e$ and then proving $c_3$ twice.
\end{itemize}
Similarly, in order to verify a Hoare triple of form
$$ \{P\} \textsf{while ($e$) \{ $c_1$ \}}; c_2 \left\{Q, \left[\vec{Q}'\right]\right\}$$
where break statements appear in some branches in the loop body $c_1$, a join condition is needed (in addition to a loop invariant) in an interactive proof, since an execution may leave the loop by a break statement or by falsifying the loop condition $e$.
In VST-A, users can choose to provide a join condition (the proof will be structural) or not to provide a join condition (the proof will be nonstructural).

Certain kinds of generalized loop invariants lead to  nonstructural proofs.
Consider a red-black tree (RBT) algorithm that reestablishes 
red-black invariants after insertion. \figref{fig:rbt} is
a straightforward textbook implementation of this algorithm~\cite{intro2algo}.
When the rotation on line \ref{code:rbt-rotation} is done, 
the loop exits immediately, since the assignment statement on line \ref{code:rbt-assignblack}
ensures that the loop condition will be evaluated to false in the next iteration.
If we were to write a loop invariant on line \ref{code:rbt-inv} before
the loop condition is checked, 
we then need to both state the RBT bottom-up fixing invariant and describe the
case in which the RBT has been fixed by the final rotation and should exit immediately.
One may instead expect a single invariant on line \ref{code:rbt-assert},
and reason separately about the control flow where the loop exits after a rotation.
Such proof strategy is unavailable in common goal-directed verifiers, in which
loop invariants are compulsory, but it is supported by VST-A.

\begin{figure}[h]
\begin{lstlisting}[language=C, mathescape=true, xleftmargin=2em,xrightmargin=1em, numbers=left, frame = {TB},escapechar=|, keepspaces=true ]
void insert_balance(struct tree *p, struct tree *root) {
    ...  // pre-processing code omitted
    /*@ Inv: ... (RBT invariant)  \/ ... (loop exit property) */ |\label{code:rbt-inv}|
    while (p != root && p->parent->color != RED) {
        /*@ Assert: ... (RBT invariant) */ |\label{code:rbt-assert}|
        struct tree *p_par = p->parent,  *p_gpar = p_par->parent;
        if (p_par == p_gpar->left) {    // p's parent is a left child
            struct tree *p_uncle = p_gpar->right;
            if (p_uncle->color == RED) {
                p_par->color = BLACK; p_uncle->color = BLACK;
                p_gpar->color = RED; p = p_gpar;
            } else {
                if (p == p_par->right) { p = p_par; left_rotate(p, root); }
                p->parent->color = BLACK; |\label{code:rbt-assignblack}| p_gpar->color = RED;
                right_rotate(p->parent->parent, root);|\label{code:rbt-rotation}|
        } } else { ... }                              // dual case where p's parent is a right child, omitted
    } }
\end{lstlisting}
    \caption{The fix-up function of red-black tree insertion}
    \label{fig:rbt}
\end{figure}

\section{Control flow splitting and soundness}
\label{sec:split_and_sound}
One of the most important components in VST-A is the verified split 
    function which reduces an entire C program's functional correctness to
    a series of straightline Hoare triples, based on the annotations.
Intuitively, this split function is a CFG-based computation (like our
    demonstration in Fig.~\ref{fig:ll-reverse-cfg}), and its soundness
    must ultimately relate to C's small-step semantics---any execution
    trace of the program statement can be decomposed into these
    separated paths.
However, we choose to prove this soundness theorem directly
    using VST's program logic \emph{(Verifiable C)}, instead of proving
    it indirectly by first showing the execution trace decomposition lemma
    that we mentioned above.
Our main consideration is: it is nontrivial to formally establish a theoretical connection between
     a program logic and an operational semantics, especially when
     a complicated program logic for a realistic programming language
     with a lot of subtleties are considered.
VST has already done that once---Verifiable C's soundness proof takes 60K lines of
    Coq definitions and proofs.
If we would choose to prove the split function sound using small-step semantics as intermediate
    proof steps, we might need to develop similar lengthy proofs.
    

Even in applications to other languages and other operational semantics,
it will be useful to build annotation verifiers on top of program logics
in the way that we present here, rather than try to prove soundness
directly from operational semantics.
Proved-sound verification tools tend to be based on similar program logics
(at least in terms of the core rules targeted by our split function:
the sequence rule, the consequence rule, etc.).  But they may be based
on quite different styles of operational semantics (e.g. imperative HOL \cite{DBLP:conf/tphol/BulwahnKHEM08}
     uses big step semantics and CompCert Clight uses small step semantics),
or they may (like VST \cite{appel14:plcc}
or Iris \cite{jung2018iris}) incorporate modal impredicativity.

Since most Hoare logic proof rules are syntax-oriented, we implement our \textsf{split} function
    through recursion on ClightA syntax tree (\S\ref{sec:split_fun}) and then we prove it sound by induction (\S\ref{sec:sound}).
In the rest part of this section, we will start from defining the Coq type of split results (\S\ref{sec:result}).

\subsection{The type of split result}
\label{sec:result}

\begin{figure}[ht]
    \begin{minipage}[b]{0.45\textwidth}
        \includegraphics[scale=0.6]{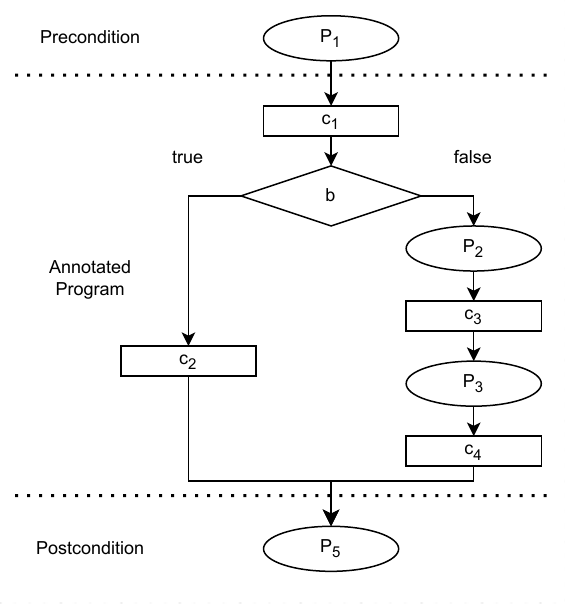}
    \end{minipage}
    \begin{minipage}[b]{0.5\textwidth}
    \begin{minipage}[b]{0.45\textwidth}
        \centering
        \includegraphics[scale=0.3]{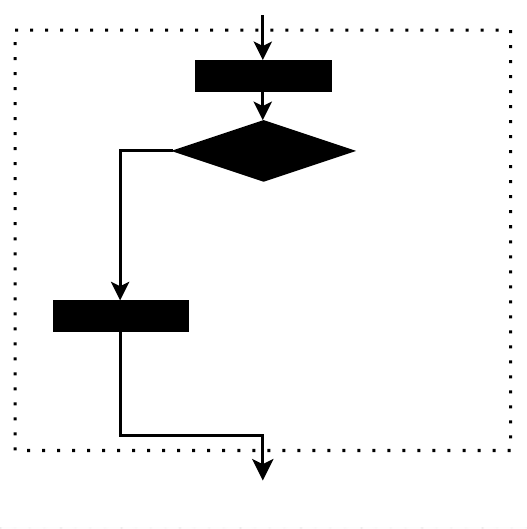} \\ \vspace{-1ex}
        \small{(Assertion-free path)}
    \end{minipage}
    \begin{minipage}[b]{0.45\textwidth}
        \centering
        \includegraphics[scale=0.3]{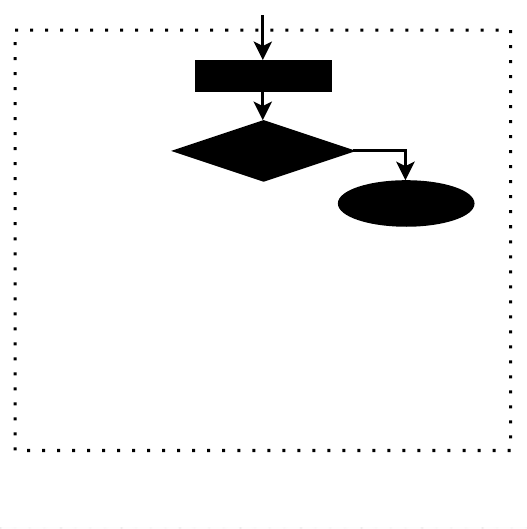} \\ \vspace{-1ex}
        \small{(Head path)}
    \end{minipage} \\ \\
    \begin{minipage}[b]{0.45\textwidth}
        \centering
        \includegraphics[scale=0.3]{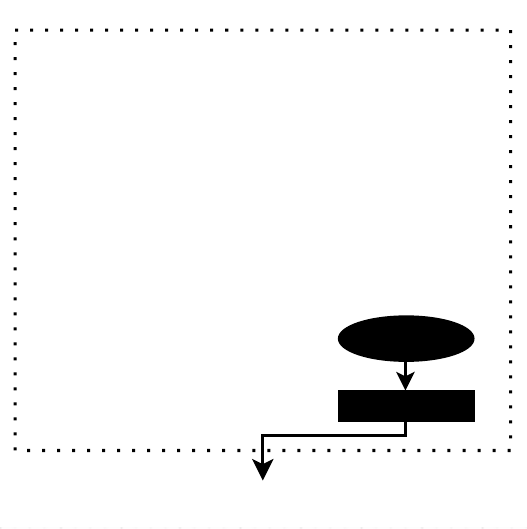} \\ \vspace{-1ex}
        \small{(Tail path)}
    \end{minipage}
    \begin{minipage}[b]{0.45\textwidth}
        \centering
        \includegraphics[scale=0.3]{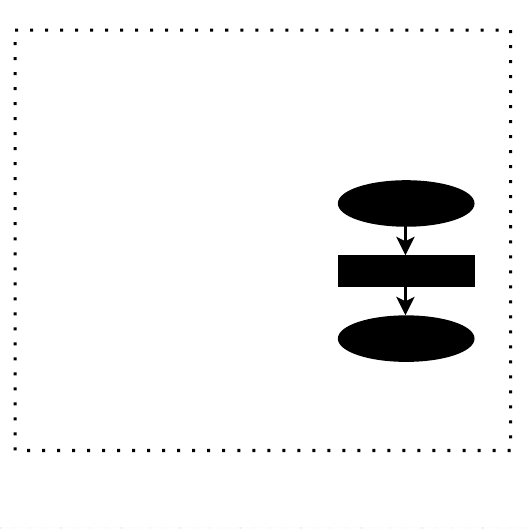} \\ \vspace{-1ex}
        \small{(Full path)}
    \end{minipage}
    \end{minipage}
\caption{Control flow graphs v.s. paths in split results}
\label{fig:type_of_paths}
\end{figure}

As illustrated in \figref{fig:type_of_paths}, control flow paths between 
    the assertions can be divided into four classes: 
\begin{enumerate}
\item head paths --- control flow paths from the precondition to an internal assertion;
\item tail paths --- control flow paths from an internal assertion to the postcondition;
\item full paths --- control flow paths between two internal assertions;
\item assertion-free paths --- control flow paths from the precondition to the postcondition (we call them \emph{assertion-free} since they pass through no assertions inside the annotated program).
\end{enumerate}

Formally, the split result is a record that consists of ``head/tail paths'', ``full paths''
and ``assertion-free paths'', which are essentially a list of \emph{basic program statements} $\vec{c_b}$
annotated with one single assertion, two assertions, and no assertions, respectively.
A basic statement can either be a primary Clight statement, $c_p$, or a special
statement, $\textsf{assume} \ e$, that represents an if-condition 
(positively or negatively) in the control flow.

Recall that in the VST program logic, a Hoare triple has multiple postconditions 
    for different kinds of program exits 
    (i.e., exit by break, by continue, by return,or normal fall-through). 
Correspondingly,
    the split result also makes distinctions among the different exits. 
Thus, in the definition of our intermediate split result, the record 
    contains one set of ``full paths'' between the annotated assertions, 
    one set of ``head paths'' from the entry point to the internal assertions, 
    four sets of ``tail paths'' from the internal assertions to the four different 
    kinds of exits, and four sets of ``assertion-free paths'' from the entry point 
    to the four different kinds of exits.
To handle existential variables in their scope, full paths can be universally quantified.
With these fields, the split result record can sufficiently
reveal all control flow information in a ClightA program.
\figref{fig:ResultAST} shows the definition.

\newcommand{\postassert}[1]{\mkern-8mu\mathrel{-\mkern-6mu\{#1\}}}
\newcommand{\preassert}[1]{\mathrel{\{#1\}\mkern-6mu-}\mkern-8mu}
\newcommand{\postassertsmall}[1]{\mkern-4mu\mathrel{-\mkern-3mu\{#1\}}}
\newcommand{\preassertsmall}[1]{\mathrel{\{#1\}\mkern-3mu-}\mkern-4mu}

\newcommand{\pathnotation}{\vdash\mkern-4mu\dashv}
\newcommand{\prenotation}{\dashv}
\newcommand{\postnotation}{\vdash}
\newcommand{\atomnotation}{-}

\newcommand{\pnormalatom}{\boldsymbol{p}_{\atomnotation}^{\text{nor}}}
\newcommand{\pbreakatom}{\boldsymbol{p}_{\atomnotation}^{\text{brk}}}
\newcommand{\pcontinueatom}{\boldsymbol{p}_{\atomnotation}^{\text{con}}}
\newcommand{\preturnatom}{\boldsymbol{p}_{\atomnotation}^{\text{ret}}}
\newcommand{\ppre}{\boldsymbol{p}_{\prenotation}}
\newcommand{\ppath}{\boldsymbol{p}_{\pathnotation}}
\newcommand{\ppost}{\boldsymbol{p}_{\postnotation}}
\newcommand{\patom}{\boldsymbol{p}_{\atomnotation}}
\newcommand{\pnormalpost}{\boldsymbol{p}_{\postnotation}^{\text{nor}}}
\newcommand{\pbreakpost}{\boldsymbol{p}_{\postnotation}^{\text{brk}}}
\newcommand{\pcontinuepost}{\boldsymbol{p}_{\postnotation}^{\text{con}}}
\newcommand{\preturnpost}{\boldsymbol{p}_{\postnotation}^{\text{ret}}}

\newcommand{\qnormalatom}{\boldsymbol{q}_{\atomnotation}^{\text{nor}}}
\newcommand{\qbreakatom}{\boldsymbol{q}_{\atomnotation}^{\text{brk}}}
\newcommand{\qcontinueatom}{\boldsymbol{q}_{\atomnotation}^{\text{con}}}
\newcommand{\qreturnatom}{\boldsymbol{q}_{\atomnotation}^{\text{ret}}}
\newcommand{\qpre}{\boldsymbol{q}_{\prenotation}}
\newcommand{\qpath}{\boldsymbol{q}_{\pathnotation}}
\newcommand{\qpost}{\boldsymbol{q}_{\postnotation}}
\newcommand{\qatom}{\boldsymbol{q}_{\atomnotation}}
\newcommand{\qnormalpost}{\boldsymbol{q}_{\postnotation}^{\text{nor}}}
\newcommand{\qbreakpost}{\boldsymbol{q}_{\postnotation}^{\text{brk}}}
\newcommand{\qcontinuepost}{\boldsymbol{q}_{\postnotation}^{\text{con}}}
\newcommand{\qreturnpost}{\boldsymbol{q}_{\postnotation}^{\text{ret}}}

\begin{figure}[h]
$$
\begin{array}{c}
\begin{array}{rrl}
    \text{Basic statement}:             &c_b := & c_p \mid \textsf{assume} \ e \\
    \text{Assertion-free paths}:                  &p_{\atomnotation} := & \vec{c_b} \\
    \text{Head paths}:    &p_{\prenotation} :=&  \vec{c_b} \postassert{P} \\
    \text{Tail paths}:   &p_{\postnotation} :=&  \preassert{P} \vec{c_b} \\
    \text{Full paths}:           &p_{\pathnotation} :=&  \preassert{P_1} \vec{c_b}  \postassert{P_2} \
                                                        \mid \  \forall (x:A).\ {p_{\pathnotation}} \\
\end{array} \\
\\
\splitfun{(C)}=
\left\{
    \begin{array}{ll}
        \pnormalatom, \pbreakatom, \pcontinueatom, \preturnatom, \\
        \pnormalpost, \pbreakpost, \pcontinuepost, \preturnpost, \\
        \ppre, \ppath
    \end{array}
\right\}, \text{ where }
\begin{array}{l}
    \pnormalatom, \pbreakatom, \pcontinueatom, \preturnatom \subseteq \text{Assertion-free paths}, \\
    \pnormalpost, \pbreakpost, \pcontinuepost, \preturnpost \subseteq \text{Tail paths}, \\
    \ppre \subseteq \text{Head paths}, ~ \ppath \subseteq \text{Full paths}
\end{array}
\end{array}
$$
    \caption{The type of split results}
    \label{fig:ResultAST}
\end{figure}


By supplementing ``head/tail paths'' or ``assertion-free paths'' 
with the pre-/postconditions, we can interpret 
the split result into a collection of 
closed Hoare triples as hypotheses of $\splitfun$'s soundness theorem (these hypotheses are illustrated in \figref{fig:type_of_paths} and formally defined in \figref{fig:hypos}):

\begin{theorem}[Soundness]
\label{thm:general_split_sound}
For any ClightA program $C$ and pre-/post-conditions $P$, $Q$, $\brkcnd{Q}$, $\concnd{Q}$ and $\retcnd{Q}$, if $\splitfun{(C)} = \left\{
\pnormalatom,
\pbreakatom,
\pcontinueatom,
\preturnatom,
\pnormalpost,
\pbreakpost,
\pcontinuepost,
\preturnpost,
\ppre,
\ppath\right\}$ and
\begin{itemize}
\item[(a)] all straightline Hoare triples from the precondition $P$ to internal assertions are provable,
\item[(b)] all straightline Hoare triples from internal assertions to the postconditions $\vec{Q}$ are provable,
\item[(c)] all straightline Hoare triples between internal assertions are provable,
\item[(d)] all straightline Hoare triples from the precondition $P$ to the postconditions $\vec{Q}$ are provable,
\end{itemize}
i.e. (defined in \figref{fig:hypos}),
\begin{itemize}
\item[(a)] $\semaxprea(P, \ppre)$,
\item[(b)] $\semaxposta(Q, \pnormalpost)$, $\semaxposta(\brkcnd{Q}, \pbreakpost)$, $\semaxposta(\concnd{Q}, \pcontinuepost)$ and $\semaxposta(\retcnd{Q}, \preturnpost)$,
\item[(c)] $\semaxpatha(\ppath)$,
\item[(d)] $\semaxatoma(P, Q, \pnormalatom)$, $\semaxatoma(P, \brkcnd{Q}, \pbreakatom)$, $\semaxatoma(P, \concnd{Q}, \pcontinueatom)$ and \\ $\semaxatoma(P, \retcnd{Q}, \preturnatom)$,
\end{itemize}
then $\hoaretriple{P}{\eraseannot{C}}{Q, \left[\brkcnd{Q}, \concnd{Q}, \retcnd{Q}\right]}$, where $\eraseannot{C}$ represents the result of erasing all annotations from $C$.
\end{theorem}

Noticing that all of the control flows in a function body should end with a return statement, we directly use the following corollary in VST-A.

\begin{corollary}
\label{thm:split_sound}
For any ClightA program $C$ and pre-/post-conditions $P$ and $Q$, if $$\splitfun{(C)} = \left\{\nil, \nil, \nil,\preturnatom,\nil,\nil,\nil, \preturnpost, \ppre, \ppath \right\}$$ and (a) $\semaxprea(P, \ppre)$, (b) $\semaxposta(Q, \preturnpost)$, (c) $\semaxpatha(\ppath)$, and (d) $\semaxatoma(P, Q, \preturnatom)$, then $\{P\}\eraseannot{C}\{Q\}$.
\end{corollary}

\paragraph{Remark}
CompCert Clight does not have an assume statement.
We choose to encode the assume statement into Clight AST, and encode straightline Hoare triples into VST Hoare triples, so that VST-A's users can directly use VST's tactics to prove those straightline triples.
Specifically,
$$ \textbf{assume} \ e \ \triangleq \textbf{if} \ (e) \ \text{skip} \ \textbf{else} \ \text{break};$$
$$\hoaretriple{P}{c_{1};c_{2};...;c_{n}}{Q} \triangleq \hoaretriple{P}{c_{1};c_{2};...;c_{n}}{Q, [\top, \bot, \bot]}.$$
We proved:
\begin{lemma}
\label{thm:assume_fact}
For straightline Hoare triples, $\hoaretriple{P}{\textbf{assume} \ e; \ c}{Q}$ if and only if $\hoaretriple{P \land e \neq 0}{c}{Q}$.
\end{lemma}

\begin{figure}[t]
$$
    \begin{array}{l}
        \semaxpre(P, \vec{c_b}\postassert{Q}) = \hoareabbr{P}{\vec{c_b}}{Q}{\top} \\
        \semaxprea(P,\ppre) = \bigwedge_{p\in\ppre} \ \semaxpre ( P , p) \\
    \\
        \semaxatom(P, Q, \vec{c_b}) = \hoareabbr{P}{\vec{c_b}}{Q}{\top} \\
        \semaxatoma(P, Q, \patom) = \bigwedge_{p\in\patom} \ \semaxatom ( P , Q,  p) \\
    \\
    \semaxpost(Q, \preassert{P}\vec{c_b})  = \hoareabbr{P}{\vec{c_b}}{Q}{\top}\\
    \semaxposta(Q, \ppost) = \bigwedge_{p\in\ppost} \ \semaxpost ( Q,  p) \\
    \\ 
    \semaxpath(\preassert{P}\vec{c_b}\postassert{Q}) = \hoareabbr{P}{\vec{c_b}}{Q}{\top} \\
    \semaxpath(\forall x.\ {p_{\pathnotation}}) = \forall x. \ \semaxpath(p_{\pathnotation}) \\
    \semaxpatha(\ppath) = \bigwedge_{p\in\ppath} \ \semaxpath ( p) \\
    \end{array}
$$
\vspace{-2ex}
    \caption{Hypotheses of the soundness theorem}
    \label{fig:hypos}
\end{figure}

\subsection{Split function}
\label{sec:split_fun}

The core \textsf{split} function is defined by recursion on the abstract syntax tree
of the input ClightA program.
Note that the \textsf{split} function is a partial function, since we will not try to compute
    the reduction result if there is an assertion-free loop in the CFG.

\paragraph{Base cases (\figref{fig:split1})}
For primary statements $c_p$, the normal assertion-free path set 
    (the $\pnormalatom$ field) is a singleton of the statement itself, 
    i.e. $\{[c_p]\}$, and all other path sets are empty.
For a $\cbreak$ statement, the assertion-free break-exit path set 
    is a singleton of an empty list of basic statements, i.e. $\{[]\}$. 
The split results of $\ccontinue$ and $\creturn$ are similar.
For the assertion annotation $\cassert{P}$,
    the split result has only one head path $\singleton{[]\postassert{P}}$
    and one normal tail path $\singleton{\preassert{P}[]}$.

\newcommand{\assumee}{\textsf{assume } e}
\newcommand{\assumenote}{\textsf{assume } \neg e}

\begin{figure}[h]
$$\begin{array}{l}
    \splitfun{(c_p)} = \left\{
        \begin{array}{ll}
            \singleton{ [c_p] }, \nil, \nil, \nil \\
            \nil,  \nil, \nil, \nil  \\
            \nil,  \nil
        \end{array}
    \right\} 
    \,
    \splitfun{(\cbreak)} = \left\{
        \begin{array}{ll}
            \nil, \singleton{ [] }, \nil, \nil  \\
            \nil,  \nil, \nil, \nil \\
            \nil,  \nil
        \end{array}
    \right\} 
    \,
    \splitfun{(\creturn)} = \left\{
        \begin{array}{ll}
            \nil, \nil, \nil, \singleton{ [] }  \\
            \nil,  \nil, \nil, \nil \\
            \nil,  \nil
        \end{array}
    \right\}
    \\
    \splitfun{(\ccontinue)} = \left\{
        \begin{array}{ll}
            \nil, \nil, \singleton{ [] }, \nil  \\
            \nil,  \nil, \nil, \nil \\
            \nil,  \nil
        \end{array}
    \right\} 
    \quad
    \splitfun{(\cassert{P})}= \left\{
        \begin{array}{ll}
            \nil, \nil, \nil, \nil   \\
            \singleton{\preassert{P}[]},  \nil, \nil, \nil \\
            \singleton{[]\postassert{P}},  \nil
        \end{array}
    \right\}
    \end{array}
    $$
    \vspace{-2ex}
    \caption{Split function for basic statements}
    \label{fig:split1}
\end{figure}

\begin{figure}[t]
    \begin{minipage}[b]{0.5\textwidth}
    \begin{center}
        \includegraphics[scale=0.5]{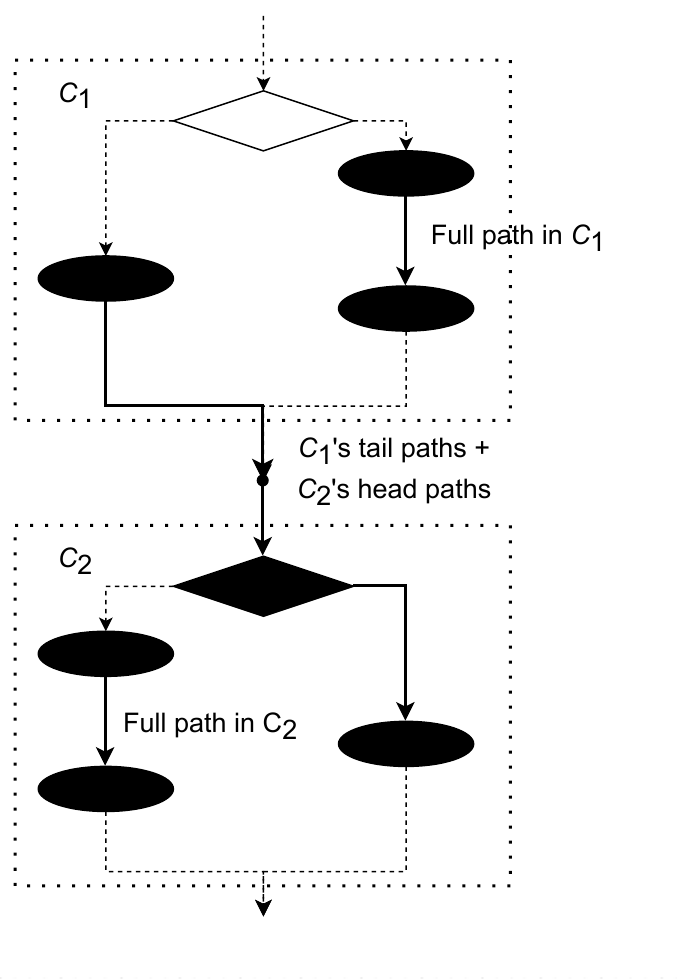} \\
        \caption{Different full paths in $C_1; C_2$}
        \label{fig:full_path_seq}
    \end{center}
    \end{minipage}
    \begin{minipage}[b]{0.45\textwidth}
        If
        $$
            \splitfun{(C_1)} = \left\{
                \begin{array}{ll}
                    \pnormalatom,
                    \pbreakatom,
                    \pcontinueatom,
                    \preturnatom, \\
                    \pnormalpost,
                    \pbreakpost,
                    \pcontinuepost,
                    \preturnpost, \\
                    \ppre,
                    \ppath
                \end{array}
            \right\}
        $$
        $$
            \splitfun{(C_2)} = \left\{
                \begin{array}{ll}
                    \qnormalatom,
                    \qbreakatom,
                    \qcontinueatom,
                    \qreturnatom, \\
                    \qnormalpost,
                    \qbreakpost,
                    \qcontinuepost,
                    \qreturnpost, \\
                    \qpre,
                    \qpath
                \end{array}
            \right\},
        $$
        then
        $$
        \begin{array}{l}
            \splitfun{(C_1 ; C_2)} = 
            \left\{
                \begin{array}{l}
                    \pnormalatom \conn \qnormalatom, \\
                    \pbreakatom \append \pnormalatom \conn \qbreakatom, \\
                    \pcontinueatom \append \pnormalatom \conn \qcontinueatom, \\
                    \preturnatom \append \pnormalatom \conn \qreturnatom, \\
                    \pnormalpost \append \pnormalpost \conn \qnormalatom, \\
                    \pbreakpost \append \qbreakpost \append \pnormalpost \conn \qbreakatom, \\
                    \pcontinuepost \append \qcontinuepost \append \pnormalpost \conn \qcontinueatom, \\
                    \preturnpost \append \qreturnpost \append \pnormalpost \conn \qreturnatom, \\
                    \ppre \append \pnormalatom \conn \qpre, \\
                    \ppath \append \qpath \append \pnormalpost \conn \qpre
                \end{array}
            \right\}.
        \end{array}
        $$
        \caption{The definition of $\splitfun{(C_1 ; C_2)}$}
        \label{fig:split_result_seq}
    \end{minipage}
\end{figure}

\paragraph{Recursion cases}
Using sequential composition as an example, all ``full paths'' between 
the assertions in $C_1 ; C_2$ can be divided into three classes (\figref{fig:full_path_seq}): 
(1) paths that completely fall in the CFG of $C_1$;
(2) paths that completely fall in the CFG of $C_2$; and
(3) paths that combine two parts, one of which is a ``tail path'' of $C_1$ and the other of which is a ``head path'' of $C_2$. 
Such ``tail paths'' and ``head paths'' of an assertion-annotated C program can also be recursively computed.
We show our complete definition in \figref{fig:split_result_seq}.

In this definition, we use $\conn$ to represent the concatenation of two paths, 
    and overload this notation to connect two sets of paths:
$$
\begin{array}{rllll}
     p_{\postnotation} \conn q_{\prenotation} = & \preassert{P} (\vec{c_b} \app \vec{c_b}') \postassert{Q}  
     & \text{where } p_{\postnotation} = \preassert{P} \vec{c_b} 
    &  \text{ and } q_{\prenotation} = \vec{c_b}' \postassert{Q}  \\
     p_{\atomnotation} \conn q_{\prenotation} = &  (\vec{c_b} \app \vec{c_b}') \postassert{Q}   
     & \text{where } p_{\atomnotation} = \vec{c_b} 
     & \text{ and } q_{\postnotation} = \vec{c_b}'\postassert{Q}   \\
     p_{\postnotation} \conn q_{\atomnotation} = & \preassert{P} (\vec{c_b} \app \vec{c_b}')
     & \text{where } p_{\postnotation} = \preassert{P} \vec{c_b} 
     & \text{ and } q_{\atomnotation} = \vec{c_b}'  \\
     p_{\atomnotation} \conn q_{\atomnotation} = &  \vec{c_b} \app \vec{c_b}'  
     & \text{where } p_{\atomnotation} = \vec{c_b} 
     & \text{ and } q_{\atomnotation} = \vec{c_b}'  \\
     \boldsymbol{p} \conn \boldsymbol{q} \ \ = & \left\{ p \conn q \mid p \in \boldsymbol{p} , \  q \in \boldsymbol{q}  \right\} 
\end{array}
$$

Computing $\splitfun{(\cifthenelse{b}{C_1}{C_2})}$ simply adds
\textsf{assume} statements to the head of all head paths and 
assertion-free paths in the two if-branches, and returns 
the union of the two split results.
Computing $\splitfun{(\cloop{C_1}{C_2})}$ is similar.
Detailed definitions can be found in \seeappendix and our Coq development.

\paragraph{Handling logical variables (\figref{fig:split_result_exgiven})} 
The focus of computing $\splitfun{(\ExGiven{x}{A}{P(x)}{C_1})}$
is to handle the logical variable $x$.
(1) The $\ExGivensymbol$ structure has an existentially quantified
    precondition $P(x)$ in the head. Therefore,
    there are no assertion-free paths in the split result,
    and the result of the head paths is a singleton of $[]\postassert{\EX{x}{A}{P(x)}}$.
(2) The tail paths in $\ExGiven{x}{A}{P(x)}{C_1}$ can either be paths from the precondition $\EX{x}{A}{P(x)}$ to exits of $C_1$ or tail paths of $C_1$ itself.
When these tail paths connect to head paths later, the postconditions of those head paths will not be in the scope of $x$.
Thus, we existentially quantify over the variable $x$ in all those tail paths' preconditions now,
   i.e. if $\preassert{Q(x)}{\vec{c_b}}$ is a tail path of $C_1$,
    then $\preassert{\EX{x}{A}{Q(x)}}{\vec{c_b}}$ is a tail path of $\ExGiven{x}{A}{P(x)}{C_1}$.
(3) In full paths, we need to unify the existential variable $x$ in $P$
        with those in the head paths that are split from $C_1$,
        so that $x$ can be shared among the pre-/post-conditions
        of each combined full path.
    Full paths in $C_1$ ($\ppath$) are also collected after adding a
        universal binder $x$ to the result.
In our Coq development, we implement this definition using Coq dependent types. We put technique details in \seeappendix.

\begin{figure}[t]
$$
\begin{array}{l}
\text{If} \
        \splitfun{(C_1)} = \left\{
            \begin{array}{ll}
                \pnormalatom,
                \pbreakatom,
                \pcontinueatom,
                \preturnatom, \\
                \pnormalpost,
                \pbreakpost,
                \pcontinuepost,
                \preturnpost, \\
                \ppre,
                \ppath
            \end{array}
        \right\}, \\
\text{then} \ \splitfun{(\ExGiven{x}{A}{P(x)}{C_1})} = \\
\ \ \ \ \left\{
        \begin{array}{l}
            \nil,
            \nil,
            \nil,
            \nil \\
            \left\{  \preassert{\EX{x}{A}{Q}} \vec{c_b} \mid \preassert{Q} \vec{c_b}  \in \pnormalpost \right\}
                \append \singleton{ \preassert{\EX{x}{A}{P(x)}}[] } \conn \pnormalatom \\
            \left\{  \preassert{\EX{x}{A}{Q}} \vec{c_b} \mid \preassert{Q} \vec{c_b}  \in \pbreakpost \right\}
            \append \singleton{ \preassert{\EX{x}{A}{P(x)}}[] } \conn \pbreakatom \\
            \left\{  \preassert{\EX{x}{A}{Q}} \vec{c_b} \mid \preassert{Q} \vec{c_b}  \in \pcontinuepost \right\}
             \append \singleton{ \preassert{\EX{x}{A}{P(x)}}[] } \conn  \pcontinueatom \\
            \left\{  \preassert{\EX{x}{A}{Q}} \vec{c_b} \mid \preassert{Q} \vec{c_b}  \in \preturnpost \right\}
            \append \singleton{ \preassert{\EX{x}{A}{P(x)}}[] } \conn  \preturnatom \\
            \singleton{ []\postassert{\EX{x}{A}{P(x)}} } , \\
            \left\{\forall x : A. \preassert{P} \vec{c_b} \postassert{Q}  \mid \vec{c_b} \postassert{Q} \in \ppre \right\} \append
            \left\{\forall x: A. \preassert{Q} \vec{c_b} \postassert{R}  \mid \preassert{Q} \vec{c_b} \postassert{R} \in \ppath \right\} 
        \end{array}
    \right\}
\end{array}
$$
\caption{The definition of $\splitfun{(\ExGiven{x}{A}{P(x)}{C_1})}$}
\label{fig:split_result_exgiven}
\end{figure}

\subsection{Proof of soundness}
\label{sec:sound}

We prove theorem \ref{thm:general_split_sound} by induction over ClightA syntax trees.
For the convenience of presentation, we use $\semaxsplit(P, \vec{Q}, {\splitfun{(C)}})$ to represent all ten hypotheses of this soudnenss theorem:
In order words, the soundness theorem says: $\semaxsplit(P, \vec{Q}, {\splitfun{(C)}})$ implies $\hoaretriple{P}{\eraseannot{C}}{\vec{Q}}$.
$$
\begin{array}{l}
    \semaxsplit
    \left(
    P, Q, \left[\brkcnd{Q}, \concnd{Q}, \retcnd{Q}\right],
    \left\{
        \begin{array}{ll}
            \pnormalatom, \pbreakatom, \pcontinueatom, \preturnatom, \\
            \pnormalpost, \pbreakpost, \pcontinuepost, \preturnpost, \\
            \ppre, \ppath
        \end{array}
    \right\}
    \right) \\
    \begin{array}{ll}
    \triangleq &
       \begin{array}{llll}
          \semaxatoma (P , Q , \pnormalatom) & \land &
          \semaxatoma ( P , \brkcnd{Q} , \pbreakatom) & \land \\
          \semaxatoma ( P , \concnd{Q} , \pcontinueatom) & \land &
          \semaxatoma ( P , \retcnd{Q} , \preturnatom) & \land 
       \end{array} \\
    &
      \begin{array}{llll}
          \semaxposta ( Q  , \pnormalpost) & \land &
          \semaxposta  ( \brkcnd{Q} , \pbreakpost) & \land \\
          \semaxposta ( \concnd{Q} , \pcontinuepost) & \land &
          \semaxposta ( \retcnd{Q} , \preturnpost) & \land 
      \end{array} \\
    & \ \ \semaxprea ( P , \ppre) \ \land \  \semaxpatha ( \ppath ).
    \end{array}
\end{array}.
$$

In this proof, only induction steps about sequential compositions, if-statements and loops are interesting.
We describe the main idea of proving $\splitfun{(C_1;C_2)}$ sound in this section. The proofs for if-statements and loops are similar and can be found in \seeappendix.

\begin{figure}[t]
\centering
\includegraphics[width=.6\textwidth]{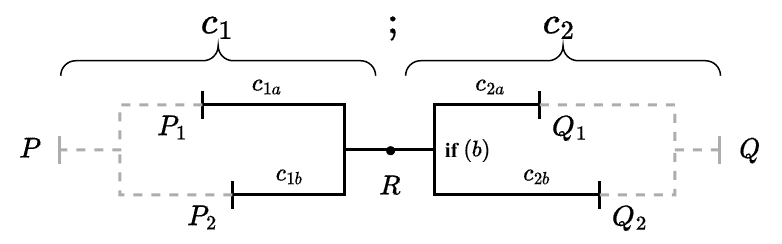}
\caption{Soundness proof example of sequential composition}
\label{fig:seq_sound}
\end{figure}

\figref{fig:seq_sound} shows an example of sequential composition
with precondition $P$, postcondition $Q$, and annotations $P_1,P_2,Q_1,Q_2$. The $\splitfun$ function will generate these straightline Hoare triples:
\begin{equation*}
\begin{array}{rcl}
    \{P_1\} & c_{1a};\textsf{assume}\ b;c_{2a} & \{Q_1\} \\
    \{P_1\} & c_{1a};\textsf{assume}\ !b;c_{2b} & \{Q_2\} \\
    \{P_2\} & c_{1b};\textsf{assume}\ b;c_{2a} & \{Q_1\} \\
    \{P_2\} & c_{1b};\textsf{assume}\ !b;c_{2b} & \{Q_2\} \\
    & \cdots &
\end{array}
\end{equation*}
In order to prove $\hoaretriple{P}{C_1\!\Downarrow\,;\, C_2\!\Downarrow}{Q}$,
we need to find an intermediate
assertion $R$ such that $\hoaretriple{P}{C_1\!\Downarrow}{R}$ and $\hoaretriple{R}{C_2\!\Downarrow}{Q}$ (according to \rulename{Semax-Seq}). These two judgments can be established by induction hypothesis and the following four Hoare triples:
\begin{equation*}
\begin{array}{cccc}
    \hoaretriple{P_1}{c_{1a}\!}{R} &
    \hoaretriple{R}{\textsf{assume}\ b;c_{2a}\!}{Q_1} &
    \hoaretriple{P_2}{c_{1b}\!}{R} &
    \hoaretriple{R}{\textsf{assume}\ !b;c_{2b}\!}{Q_2}
\end{array}
\end{equation*}
These requirements can be simply satisfied if we let $R$ be
$$\text{wp}(\textsf{assume}\ b;c_{2a}, \ Q_1) \wedge \text{wp}(\textsf{assume}\ !b;c_{2b}, \ Q_2).$$

In the general case, 
we instantiate $R$ to the conjunction 
of the weakest preconditions of all head paths and assertion-free paths in $\splitfun(C_2)$.
Using VST's higher order logic, this middle condition can be written as:
\begin{equation*}
    R \triangleq \left\{\begin{array}{l}
    \exists R. \ R \land \semaxprea(R, \qpre) \\
    \quad\begin{array}{l}
        \land ~ \semaxatoma(R, Q, \qnormalatom) \\
        \land ~ \semaxatoma(R, \brkcnd{Q}, \qbreakatom) \\
        \land ~ \semaxatoma(R, \concnd{Q}, \qcontinueatom) \\
        \land ~ \semaxatoma(R, \retcnd{Q}, \qreturnatom)
    \end{array}
\end{array}\right\}
\end{equation*}
\makeatletter

\makeatletter
\newcounter{dummy}
\newcommand\myitem[1][]{\item[#1]\refstepcounter{dummy}\def\@currentlabel{#1}}
\makeatother

According to the induction hypothesis, it suffices to prove the following two propositions.
\begin{eqnarray}
    \semaxsplit(P, R, [\brkcnd{Q}, \concnd{Q}, \retcnd{Q}], \splitfun{(C_1)})\label{eqn:seq-IH1}  \\
    \semaxsplit(R, Q, [\brkcnd{Q}, \concnd{Q}, \retcnd{Q}], \splitfun{(C_2)})\label{eqn:seq-IH2}
\end{eqnarray}
The proof of (\ref{eqn:seq-IH2}) is simple, which can be justified by the following lemma in VST.

\begin{lemma}
    For any program $c$ and postcondition $\vec{Q}$, 
    $\hoaretriple{\EX{P}{\textsf{assert}}{P \land \left(\hoaretriple{P}{c}{\vec{Q}}\right)}}{c}{\vec{Q}}$ holds.
\end{lemma}

For proposition \ref{eqn:seq-IH1}, the inversion lemma for sequencing (Lemma \ref{lemma:semax_seq_inv})
has already shown that the weakest precondition of the second statement can
serve as the intermediate assertion for the sequential composition.
Based on Lemma \ref{lemma:semax_seq_inv}, we can prove  a corresponding inversion lemma 
on the $\conn$ operator for each type of path
in the split results.

\begin{proposition}[Inversion lemmas for split results]
    \label{prop:semax_seq_inv_split}
    \,
    \begin{enumerate}
        \item If $\semaxatom(P,Q, p_{\atomnotation} \conn  q_{\atomnotation})$, then 
            $\semaxatom(P, \EXabbr{R}{R \land \semaxatom(R,Q, q_{\atomnotation})}, p_{\atomnotation})$
        \item \label{enum:semaxpre_inv} If $\semaxpre(P, p_{\atomnotation} \conn q_{\prenotation} )$, then
            $\semaxatom(P, \EXabbr{R}{R \land \semaxpre(R,q_{\prenotation})}, p_{\atomnotation})$
        \item \label{enum:semaxpost_inv} If $\semaxpost(Q, p_{\postnotation} \conn q_{\atomnotation})$, then
            $\semaxpost(Q, \EXabbr{R}{R \land \semaxatom(R,Q,q_{\atomnotation})}, p_{\postnotation})$
        \item If $\semaxpath(p_{\postnotation} \conn q_{\prenotation})$, then
            $\semaxpost(\EXabbr{R}{R \land \semaxpre(R,q_{\prenotation})}, p_{\postnotation})$
    \end{enumerate}
\end{proposition}


According to the conjunction rule, 
$\semaxatom$ and $\semaxpost$ from
Proposition \ref{prop:semax_seq_inv_split}
still hold if we combine all of those weakest preconditions
of $C_2$'s partial paths (i.e. those preconditions are 
such that $\semaxatom(R,Q,q_{\atomnotation})$ and
$\semaxpre(R,q_{\prenotation})$). 
Formally, Proposition \ref{prop:semax_seq_inv_split} can be 
    extended into the following form by the conjunction rule.


\begin{proposition}[Grouped Inversion lemmas]
    \label{prop:semax_seq_inv_split_group}
    \,
    \begin{enumerate}
        \item If $\semaxatoma(P,Q, \patom \conn  \qatom)$, then 
            $\semaxatoma(P, \EXabbr{R}{R \land \semaxatoma(R,Q, \qatom)}, \patom)$
        \item If $\semaxprea(P, \patom \conn \qpre )$, then 
            $\semaxatoma(P, \EXabbr{R}{R \land \semaxprea(R,\qpre)}, \patom)$
        \item \label{enum:semaxpost_inv_group} If $\semaxposta(Q, \ppost \conn \qatom)$, then 
            $\semaxposta(Q, \EXabbr{R}{R \land \semaxatoma(R,Q,\qatom)}, \ppost)$
        \item If $\semaxpatha(\ppost \conn \qpre)$, then
        $\semaxposta(\EXabbr{R}{R \land \semaxprea(R,\qpre)}, \ppost)$
    \end{enumerate}
\end{proposition}

\begin{theorem}[Conjunction Rule]
    \label{thm:conj_rule}
    If Hoare triples $\hoareabbr{P}{c}{Q_1} {Q^{\prime}_{1}}$ and 
    $\hoareabbr{P}{c}{Q_2}{Q^{\prime}_{2}}$ are derivable, then 
    ${\{P\}\,c\,\left\{Q_1 \wedge Q_2, [\vec{Q^{\prime}_{1}} 
    \wedge \vec{Q^{\prime}_{2}}]\right\}}$ is derivable. 
\end{theorem}

Next, we also use the conjunction rule to combine the
weakest preconditions of the different kinds of paths
and prove proposition \ref{eqn:seq-IH1},
which completes the soundness proof of $\splitfun{(C_1;C_2)}$.

However, the conjunction rule is not ubiquitous among the Hoare logic variants proposed
in the literature: for example, the current VST program logic cannot derive
the conjunction rule.
We will leave the discussion of the conjunction rule to \S\ref{sec:conj}. 
For now, we assume that the conjunction rule holds,
so that Proposition \ref{prop:semax_seq_inv_split_group}
and the proof of  $\splitfun{(C_1;C_2)}$ soundness can hold.

\section{Conjunction rule and preciseness}
\label{sec:conj}

The conjunction rule is natural in traditional Hoare logics and separation logics
    for sequential programs, but some extensions to the logics will make the conjunction
    rule inadmissible. 
In this section, we more extensively discuss why the conjunction rule is 
    required by our soundness proof (\S\ref{sec:why_by_reduction}, \S\ref{sec:why_by_counterexample}).
To make the conjunction rule admissible in VST-A (\S\ref{sec:conj_all}, \S\ref{sec:conj_store}, \S\ref{sec:conj_call}), we identify a new notion of 
    preciseness to restrict the function specifications being called during verification.
We also discuss the trade-offs of using conjunction rules and precise
    function specifications, and we suggest some future directions for improvement (\S\ref{sec:conj_remark}).

\subsection{A small example}
\label{sec:why_by_reduction}

Suppose we would like to prove the Hoare triple
\begin{equation}
\hoaretriple{P}{c_1; \ \textbf{if} \ (b) \ c_2 \ \textbf{else} \ c_3}{ R } \label{eqn:tgt}
\end{equation}
given that the following split results hold (here, we assume that $c_1$, $c_2$ and $c_3$ are primary statements):
\begin{eqnarray}
& \hoaretriple{P}{c_1; \ \textbf{assume} \ b; \ c_2 }{ R } \label{eqn:path1} \\
& \hoaretriple{P}{c_1; \ \textbf{assume} \ !b; \ c_3 }{ R } \label{eqn:path2}
\end{eqnarray}
By inversion on sequential composition (lemma \ref{lemma:semax_seq_inv}), proposition (\ref{eqn:tgt}), (\ref{eqn:path1}) and  (\ref{eqn:path2}) are equivalent to:
\begin{eqnarray*}
&& \hoaretriple{P}{c_1}{\text{wp}(\textbf{if} \ (b) \ c_2 \ \textbf{else} \ c_3, \ R) } \\
&& \hoaretriple{P}{c_1}{\text{wp}(\textbf{assume} \ b; \ c_2, \ R) } \\
&& \hoaretriple{P}{c_1}{\text{wp}(\textbf{assume} \ !b; \ c_3, \ R) }
\end{eqnarray*}
Furthermore, by inversion (Lemmas~\ref{lemma:semax_seq_inv} and \ref{lemma:inversion_if}) and properties of \textbf{assume} (Lemma~\ref{thm:assume_fact}), $\text{wp}(\textbf{if} \ (b) \ c_2 \ \textbf{else} \ c_3, \ R)$ is equivalent to
$$\text{wp}(\textbf{assume} \ b; \ c_2, \ R) \ \land \text{wp}(\textbf{assume} \ !b; \ c_3, \ R)$$ 
Thus, the split function's soundness on the example above can be reduced to an instance of the conjunction rule.

\subsection{Unsoundness when the conjunction rule is inadmissible}
\label{sec:why_by_counterexample}

So far, we have seen a tight connection between the conjunction rule and the split function's soundness --- our soundness proof uses the conjunction rule (\S\ref{sec:sound}) and a very simple instance of this soundness theorem can be reduced to an instance of the conjunction rule (\S\ref{sec:why_by_reduction}).
But what will happen to \textsf{split}'s soundness if the 
conjunction rule is not admissible?
Consider Hoare logic with ghost updates \cite{Iris2} as an example.
Ghost states are ``logical states'' that help with the program's proof,
and they particularly useful for verifying concurrent programs.\footnote{Ghost states are not the same as the ``ghost variables'' of traditional Hoare logics.  
Ghost variables are logical variables that were introduced to relate old
values of variables to current values, and to relate current values to
abstract mathematical values.  VST and Iris support ghost variables using
ordinary Coq variables; those ghost variables are fully compatible with
our VST-A program decomposition, to support data abstraction and modular verification~\cite{VSU}. Examples of such variables
in \figref{fig:ll-reverse} are $l,l_1,u,x,l_2$. }
When users prove programs with ghost states, they can
apply a ghost update when they use the consequence rule, see \rulename{Semax-Conseq-Ghost} below.
Here, $\bupdd P$ says: there is at least one possible ghost update which makes the state satisfy $P$.
\begin{mathpar}
    \inferrule*[left=Semax-Conseq-Ghost]{
         P_1 \entails \bupdd P_2 \and
         R_2 \entails \bupdd R_1 \and
         \vec{R'_2} \entails \bupdd \vec{R'_1} \and
        \hoareabbr{P_2}{c}{R_2}{R'_2}
    } {
        \hoareabbr{P_1}{c}{R_1}{R'_1}
    }
\end{mathpar}

The conjunction rule is not admissible in this logic --- 
if the proofs of $\hoaretriple{P}{c}{Q}$ and $\hoaretriple{P}{c}{Q'}$
use different and conflicting ghost updates, 
$\hoaretriple{P}{c}{Q \land Q'}$ cannot be valid since 
two conflicting ghost updates cannot happen simultaneously.

In this logic, our split function is unsound.
This loss of soundness is not determined by the way 
    we prove soundness in \S\ref{sec:sound} but by the 
    framework we propose to first split the program into individual 
    paths, which are then verified separately.
Consider the following annoated program:
\begin{lstlisting}[language=C, mathescape=true, xleftmargin=2em,xrightmargin=1em, numbers=left, frame = {TB}, escapechar=|,]
    /*@ Assert $g \mapsto A$ */
    f();
    x = nondetermined_0_or_1();
    if (x) {
        /*@ Assert $g \mapsto B_1 \land \cvar{x} = 1$ */
        f1();
    } else {
        /*@ Assert $g \mapsto B_0 \land \cvar{x} = 0$ */
        f0();
    }
    /*@ Assert $g \mapsto C_1 \land \cvar{x} = 1 \vee g \mapsto C_0 \land \cvar{x} = 0$ */
\end{lstlisting}
in which $g$ is a ghost location for storing the status of the following STS (state transition system) \cite{DBLP:conf/popl/TuronTABD13}:
$A \to A_0, \ A \to A_1, \ A_0 \to B_0, \ A_1 \to B_1, \ B_0 \to C_0, B_1 \to C_1$.
We assume that \lstinline{f()}, \lstinline{f0()}, and \lstinline{f1()}'s specifications are:
\begin{eqnarray*}
&&\text{for any } b \in \textbf{bool}, \\
&& \ \ \hoaretriple{g \mapsto (\text{if} \ b \ \text{then} \ A_1 \ \text{else} \ A_0)}{f()}{g \mapsto (\text{if} \ b \ \text{then} \ B_1 \ \text{else} \ B_0)} \\
&&\hoaretriple{g \mapsto B_0}{f0()}{g \mapsto C_0} \\
&&\hoaretriple{g \mapsto B_1}{f1()}{g \mapsto C_1}.
\end{eqnarray*}
In this example, all straightline Hoare triples in $\splitfun$'s result are provable, especially the following two triples about \lstinline{f()}:
\begin{lstlisting}[mathescape=true]
    $\{g \mapsto A\}$ f(); x = nondetermined_0_or_1(); $\textbf{assume}$ x; $\{g \mapsto B_1 \land \cvar{x} = 1 \}$
    $\{g \mapsto A\}$ f(); x = nondetermined_0_or_1(); $\textbf{assume}$ !x; $\{g \mapsto B_0 \land \cvar{x} = 0 \}$
\end{lstlisting}
For the first triple, we can choose to take the $A \to A_1$ step in ths STS before calling \lstinline{f()}. 
For the second triple, we can choose to take the $A \to A_0$ step in ths STS before calling \lstinline{f()}.
However, the whole Hoare triple is not provable since we cannot determine the value of $\cvar{x}$ before \lstinline{x = nondetermined_0_or_1()} is executed.
That breaks $\splitfun$'s soundness.

\subsection{VST-A's design choice and proof strategy}
\label{sec:conj_all}

In the current design of VST-A, we focus on sequential program verification
    and disallow all ghost updates.
VST-A uses a more restricted variant of the VST program logic.
This variant is still proved sound w.r.t. CompCert Clight semantics and the most significant change is that the ghost update operator is removed
    from the consequence rule.

Despite the removal of ghost updates, users are still able to write unrestricted higher-order 
    predicates and prove many complex sequential
    programs in VST-A.
We derive the conjunction rule (theorem \ref{thm:conj_rule}) from this stronger logic by induction over Clight abstract syntax tree.
Our inductive proof steps are all Hoare-logic-based, and we believe that such a proof strategy is (1) easier to formalize in Coq, and (2) in fact more general than a semantic-based proof, since
    it is independent of how the soundness of the Hoare logic was proved w.r.t. its semantic model.

Consider the induction step for $c = c_1;c_2$. By applying Lemma \ref{lemma:semax_seq_inv} to the
premises we obtain the following: 
$$
\begin{array}{c}
    \hoareabbr{P}{c_1}{\EXabbr{R_1}{R_1 \land \hoareabbr{R_1}{c_2}{Q_1}{Q'_1}}}{Q'_1}\\
    \hoareabbr{P}{c_1}{\EXabbr{R_2}{R_2 \land \hoareabbr{R_2}{c_2}{Q_2}{Q'_2}}}{Q'_2}
\end{array}
$$
We can apply the induction hypothesis of $c_1$ and make use of
\textsc{Semax-Conseq} to obtain the following:
$$
\{P\}{c_1}\left\{
    \begin{array}{l}
        \exists R. \ R
        \land \hoareabbr{R}{c_2}{Q_1}{Q'_1} \\
        \ \ \ \ \ \ \ \ \ \
        \land \hoareabbr{R}{c_2}{Q_2}{Q'_2}
    \end{array},
    \left[\vec{Q'_1}\land\vec{Q'_2}\right]
    \right\}
$$
where $R$ can be instantiated as $R_1 \land R_2$. According to
\textsc{Semax-Seq}, we are left to prove:
$$
\left\{
    \begin{array}{l}
        \exists R. \ R
        \land \hoareabbr{R}{c_2}{Q_1}{Q'_1} \\
        \ \ \ \ \ \ \ \ \ \ 
        \land \hoareabbr{R}{c_2}{Q_2}{Q'_2}
    \end{array}
\right\} {c_2}
\left\{ 
    \begin{array}{l}
        Q_1 \land Q_2, \\
        {\left[\vec{Q'_1} \land \vec{Q'_2}\right]}         
    \end{array}
\right\}
$$
Using \textsc{Extract-Exists} and \textsc{Extract-Prop}, it suffices to prove:
$$\hoareabbr{R}{c_2}{Q_1}{Q'_1}, \hoareabbr{R}{c_2}{Q_2}{Q'_2} \ \Rightarrow \ 
\left\{R\right\} {c_2}
\left\{ Q_1 \land Q_2, \left[\vec{Q'_1} \land \vec{Q'_2}\right] \right\}
$$
which immediate follows the induction hypothesis
of $c_2$.

For other induction steps (e.g. for if-statements and for loops), the proof idea is somewhat similar in that
their corresponding inversion lemmas are first applied, then the induction hypotheses can be used to complete the proof.
For control flow statements (break, continue and return), the proof is trivial.
Thus, we are only left to derive the conjunction rules
for primary statements.

\subsection{The conjunction rule for memory stores}
\label{sec:conj_store}

Consider a simple store statement and corresponding proof rule in VST.

\begin{equation}
    \label{eqn:store-spec}
    \forall \ v.
    \begin{array}{c}
        \left\{ \begin{array}{l}
            p \mapsto v \sepcon (p \mapsto v' \wand R)
        \end{array}\right\} \
        * p = v' \
        \left\{ R \right\} \
    \end{array}
\end{equation}

The precondition states that the heap can be
    split into two parts. 
One part is a singleton heap, in which the
    nonaddressable variable $p$ is a pointer to 
    the old value $v$, 
    and it is described by the mapsto predicate $x \mapsto v$. 
The other part should satisfy the postcondition $R$
    when joined with a singleton heap that stores the new value $v'$.

Note that in this specification, the old value $v$ is universally quantified. 
If we want to conjoin the postconditions of
    two Hoare triples about $*p = v'$, we need to unify the two
    different instantiations of $v$ into one.
To be specific, the following property needs to hold:

\begin{proposition}[Preciseness of store]
    \label{thm:write_conj}
    \begin{displaymath}
        \begin{array}{rl}
            & (\exists \ v_1. \ p \mapsto v_1 \sepcon (p \mapsto v' \wand R_1)) \ \land \\
            & (\exists \ v_2. \ p \mapsto v_2 \sepcon( p \mapsto v' \wand R_2)) \\
            \entails & \exists \ v. \  p \mapsto v \sepcon  (p \mapsto v' \wand R_1 \land R_2)
        \end{array}
    \end{displaymath}
\end{proposition}

In VST-A, we prove this through
semantic-model-level reasoning based on its underlying memory model.
Then, the conjunction rule for such simplified store statements can be logically derived.
We leave the detailed proof for \seeappendix.

In VST, assignment statements $ \cassign{e_1}{e_2}$ include
   (1)~assigning a value to a nonaddressable variable,
   (2)~loading a value from memory to a nonaddressable variable
   and (3)~storing a value in memory.
VST's higher-order assertion language also supports separation logic predicates with fractional permissions.
In VST-A, we have proved that the conjunction rule holds 
    for all primary set/load/store operations.

\subsection{The conjunction rule for function calls}
\label{sec:conj_call}
The store rule mentioned above can be treated as a 
specification for a function call that contains only a single 
store instruction. We have shown that Proposition \ref{thm:write_conj}
is an important property for deriving the conjunction rule. In this section, we
generalize this property, which we refer to as \emph{precise function specifications},
to derive the conjunction rule for function calls.

As previously mentioned, VST-A function specification has a form of
$$\funspecabbr{\vec{a}:\vec{A}}{\lambda \vec{y}. P}{\lambda r. Q}, $$ 
where $P$ is a precondition parameterized by a list of formal 
parameters $\vec{y}$, $Q$ is a postcondition parameterized 
by the return value $r$, and types $\vec{A}$
represents the type of logical values to be shared between 
$P$ and $Q$. Therefore, both $P$ and $Q$ will also be abstracted 
over a set of logical variables $\vec{a}$ typed $\vec{A}$.
We define the precise function specifications as follows:

\begin{definition}[Precise function specification]
    \label{def:precise_fun}
    $\funspecabbr{\vec{a}:\vec{A}}{\lambda \vec{y}. P}{\lambda r. Q}$
    is precise, if for any formal parameters $\vec{b}$, return value $r$ and
    assertions $R_1, R_2$, it holds that
    \begin{displaymath}
        \begin{array}{rl}
            & \left(\exists \vec{x_1}: \vec{A}. \ P \ \vec{b} \ \vec{x_1} \sepcon \left( Q \ r \ \vec{x_1} \wand R_1 \right)\right)
            \land \left(\exists \vec{x_2}: \vec{A}. \ P \ \vec{b} \ \vec{x_2} \sepcon \left( Q \ r \ \vec{x_2} \wand R_2 \right)\right) \\
            \entails & \exists \vec{x}: \vec{A}. \ P \ \vec{b} \ \vec{x} \sepcon ( Q \ r \ \vec{x} \wand R_1 \land R_2)
        \end{array}
    \end{displaymath}
\end{definition}

In VST-A's program logic (a variant of VST's program logic), the function call rule, \rulename{Semax-Call}, requires the callee function specification to be precise.

$$
\inferrule*[left=Semax-Call]{
    \phi = \funspecabbr{\vec{x}:\vec{A}}{\lambda \vec{y}. P}{\lambda r. Q} \\
    f \text{ satisfies } \phi \and \text{$\phi$ is a precise specification}
} {
     \hoareabbr
        { {\vec{e}}\hasvalue{\vec{b}} \land
            \exists \vec{x}:\vec{A}.\  P \, \vec{b} \, \vec{x} \sepcon
                \left( Q \, \textsf{a} \, \vec{x} \wand R \right) }
        { \cassign{\textsf{a}}{f(\vec{e})}}
        {R}{\bot}
}
$$

The notion of ``precise function specification'' that 
    we propose is defined with respect to an operation
    (e.g. a store statement or a function call), 
    while traditionally, ``preciseness''
    is defined for a predicate. 
A typical example of a precise predicate
    is the $p \mapsto v$ predicate used in \rulename{Semax-Store}.

\begin{definition}[Precise predicate]
    A predicate $P$ is precise if for any $Q_1, Q_2$,
    $$(P \sepcon Q_1) \land (P \sepcon Q_2) = P \sepcon (Q_1 \land Q_2)$$
\end{definition}

In fact, our definition of precise function specifications includes
    a common use case of precise predicates.
As proved by \citet{Gotsman2011} and \citet{Vafeiadis2011}, if the conjunction
    rule is expected in concurrent separation logic with locks,
    the resource invariant for locks should be a precise predicate.
Consider the following specification for a release operation for lock $l$:
$$
\forall \ l \ \pi. \ \left\{ \pi \text{ is readable} \land \textsf{locked}_{\pi}(l, R) \sepcon R \right\}
\textsf{release}(l) \left\{ \textsf{unlocked}_{\pi}(l, R) \right\}
$$
where $R$ is the predicate that describes the locked memory. We use $\textsf{locked}_\pi$
and $\textsf{unlocked}_\pi$ to denote whether this memory is owned by the current thread or not; $\pi$ is a fractional permission.
It is clear that if the resource invariant $R$ is a precise predicate, then
one can show that this specification is a precise specification.
The precise function specification also makes a difference in the sequential setting
  we are considering. 
One example of a non-precise function specification is:
\begin{lstlisting}[language=C, mathescape=true, xleftmargin=2em,xrightmargin=1em, numbers=left, frame = {TB}, escapechar=|,]
void foo (int * p)
  /*@ $\With \ q  \quad  \Require \ \cvar{p} \mapsto 0 \sepcon F(q) \quad  \Ensure  \ \cvar{p} \mapsto 1 \sepcon F(q)$ */
  { * p = 1; return; }      
\end{lstlisting}
Here $F$ is a predicate dependent on a logical variable $q$ describing the frames that the function $\cvar{foo}$ does not modify. It is possible to find such $F$ that holds for two different instantiations of $q$, and breaks the precise function specification requirement.

In our Coq development, we find that, for most function specifications, we can derive that for any formal parameters $\vec{b}$, logical variables $\vec{x_1}, \vec{x_2}$:
\begin{equation}
    \left(P \ \vec{b} \ \vec{x_1} \sepcon \TT\right)
    \land \left(P \ \vec{b} \ \vec{x_2} \sepcon \TT\right)
    \entails \vec{x_1} = \vec{x_2}
    \label{eqn:precise_equal}
\end{equation}
Then, to prove a specification precise, it is sufficient to show that $P$ is a precise predicate. A typical example is the store predicate ``$\mapsto$''. Clearly:
\begin{equation}
    \forall\ p\ v_1\ v_2.\
    \left((\cvar{p}\mapsto v_1 * \TT) \land (\cvar{p}\mapsto v_2 * \TT)\right)\entails v_1=v_2
    \label{eqn:precise_equal_mapsto}
\end{equation}
For linked lists' representation predicate $\ll$, we also have:
\begin{equation}
    \forall\ p\ l_1\ l_2.\ 
    ((\ll\left(\cvar{p},l_1\right) * \TT) \land (\ll\left(\cvar{p},l_2\right) * \TT))\entails l_1=l_2
    \label{eqn:precise_equal_ll}
\end{equation}

Moreoever, property~(\ref{eqn:precise_equal}) is composable. For example, in order to prove:
\begin{displaymath}
    ((\cvar{p}\mapsto x_1 * \ll(x_1,l_1)) * \TT)
    \land
    ((\cvar{p}\mapsto x_2 * \ll(x_2,l_2)) * \TT)
    \entails x_1 = x_2 \land l_1 = l_2
\end{displaymath}
we can first use (\ref{eqn:precise_equal_mapsto}) to derive $x_1=x_2$, then use after (\ref{eqn:precise_equal_ll}) to derive $l_1=l_2$.

It is well-known that precise predicates are also composable, thus we developed several extensible automatic tactics in VST-A to help users prove the precise specification by combining property~(\ref{eqn:precise_equal}) and predicates' preciseness.%
\footnote{In several cases only weaker equivalence relations can be derived from the conjunction. Our tactics also support proof automation for that, if this relation implies the equality of the specification}

We believe that most C functions have specifications that are naturally precise. Non-precise specifications are usually caused by separation logic conjuncts that describe inaccessible memory slices to the function, such as the $F(q)$ proposition in the example above. In practical verification tasks, one can usually remove these conjuncts using the ``frame rule'' and obtain a precise specification. The pruned part of the specification can be expressed elsewhere in the program, where the memory slice is really used.

\subsection{Discussion and future work}
\label{sec:conj_remark}
In our current design of VST-A, we require the conjunction rule to be
    derivable in the logic to ensure the splitting algorithm's soundness.
We do not consider our current design choice
    as a fatal problem for future extensions.
Here are some potential research directions that may
    support ghost updates or remove the restriction of only using precise 
    specifications in the future.

\paragraph{Verify the sequential fragments of a concurrent program}
As was discussed in \S\ref{sec:why_by_counterexample}, the existence
of the conjunction rule forbids the use of ghost updates. However, there
are ways to verify the sequential fragment of a concurrent program
with VST-A. We have proved the following property:

\begin{theorem}
    If $\hoareabbrr{P}{c_1;c_2}{R}{R'}$, where $\,_{\bupd}$ indicates that
    the triple is derivable from the original VST logic that
    supports ghost state updates,
    then there exists $Q,\vec{Q'}$ such that
    $\hoareabbr{P}{c_1}{Q}{Q'}$ and $\hoareabbrr{Q}{c_2}{R}{R'}$.
\end{theorem}
The triple about $c_1$ is derivable from the stronger program logic in VST-A.
We can apply the current VST-A framework to the sequential fragment $c_1$,
and leave the rest of the program to be
verified by the full-power VST interactive verifier.

\paragraph{New annotations for ghost actions} Another possible direction is
to improve the capability of annotations in terms of expressing proofs.
For example, in future versions of VST-A, we could allow users to write either
explicit ghost commands in their programs, or write two 
consecutive assertions where the latter can be derived from the former
with a ghost update. As a result, we do not need to worry about the possibility of having two conflicting ghost updates simultaneously.


\paragraph{Additional annotations for function calls} As was discussed in \S\ref{sec:conj_call}, in order to derive the conjunction rule for function calls, the callee should meet the requirement of precise function specification. However, this requirement could be removed if the logical variables of the specification are explicitly instantiated. This avoids the possibility that proofs of multiple straightline Hoare triples containing the same function call could actually instantiate those logical variables differently, causing issues in proving the conjunction rule. Intuitively, this is like putting two assertions before and after the function call statement as pseudo "join points" in the control flow, ensuring that every function call would only appear in one unique path in the split result. Then, $\splitfun$'s soundness would be trivial for function calls.

\paragraph{Prophecy variables} In our counterexample in \S\ref{sec:why_by_counterexample}, the verification target is unprovable but will become provable if we are allowed to insert a prophecy variable \cite{DBLP:conf/lics/AbadiL88, DBLP:journals/pacmpl/JungLPRTDJ20} into the program.
In general, we may be able to prove $\splitfun$'s soundness with the help of prophecy variables even if ghost updates are permitted in the logic.

\section{Evaluation}
\label{sec:eval}

In this section we evaluate VST-A in various aspects, including verification effort, verification time, and statistics of our development. 
We also draw a comparison between VST-A and VST.

\paragraph{Verification effort using VST-A}

We test VST-A using several sets of programs and present the 
    statistics about verification effort in
    Table~\ref{tbl:eval}.
VST-A proofs are divided into two parts: 
    the annotated programs
    (which describe the main idea 
    of the proof in a way easier to read than VST proofs)
    and Coq verification of straightline 
    Hoare triples (similar to corresponding parts of VST proofs).
Therefore, we count the lines of annotations and proofs 
    separately, in the Specification and Assertion columns
    and the Proof column, respectively.

Our benchmarks include some measured by
\citeN{refinedC}, 
    including linked lists and binary search trees.
We also include a slightly larger example\textemdash 
    a small-step interpreter of a toy imperative 
    language, which includes a number of if-branches.
It can be seen from our evaluation that without
    providing additional assertions in the program code, 
    VST-A is still able to split and verify the program.

\begin{table}[htb]
\centering
\begin{tabular}{lcc|ccc|c|}
       &    &   & \multicolumn{3}{c|}{VST-A} & VST \\ 
Program & \makebox[2em][r]{Functions} & Code & Spec & Assert & Proof & Proof \\
\midrule
Basics & ~8 & ~78 & 56 & ~5 & ~84 & ~156 \\
Singly linked list & 18 & 350 & 85 & 55 & 969 & 1212 \\
Doubly linked list & ~4 & ~95 & 16 & 16 & 171 & ~213 \\
Binary search tree & ~4 & 115 & 44 & 23 & 202 & ~364 \\
Interpreter & ~5 & 337 & 47 & ~0 & 423 & ~653 \\
\bottomrule
\end{tabular}
\caption{Number of C functions, lines of C code (without annotations), VST-A specification lines annotated in the C comments, assertion annotations in the C comments, Coq proofs in VST-A, versus Coq specifications and proofs in VST.}
\label{tbl:eval}
\end{table}

We also conduct a comparison in 
    the proof effort between VST-A and VST.
The last column in Table \ref{tbl:eval} shows the lines of proofs 
    for verifying the same function in VST; compare this
    to the ``Specification+Assertion+VST-A Proof'' columns. 
The proof lines do not count auxiliary predicate definitions and 
    lemmas since they are the same in VST and VST-A. 

Not manifest in the line-count, but still important, is that the VST user must learn to use
several different tactics for different forms of control flow, and each of these has several
options depending on which assertions are supplied (e.g., if-postcondition, for-loop
continue assertion, for-loop break assertion, etc.).  In contrast, VST-A's control-flow
splitting takes care of all of this, leaving the user to learn only the
straight-line \emph{forward} and \emph{forward\_call} tactics.
In some cases, it makes Coq proof easier to automate. For example, verifying the interpreter example in Table \ref{tbl:eval} using VST-A only needs to use \emph{forward} to handle \textsf{assume}s repeatedly, but this proof strategy cannot be used in the corresponding VST proof.


To summarize, we believe that to verify the same program, the manual effort of VST-A is no more than that of VST, while in the meantime VST-A can provide more intuitive and readable proofs with annotations in the C source program.

\paragraph{Verification time for different phases in VST-A}

\begin{table}[htb]
\centering
\begin{tabular}{lcccccc}
\toprule
Program 
& Reduction
& Common
& Avg. compile
& Avg. verify
& Max. verify
& VST verify \\
\midrule
Basics             & 0.239 & 12.0 & 1.7 $\times$ 14 & ~4.8 $\times$ 14 & ~8.1 & ~29.1 \\
SLL                & 0.087 & 11.0 & 1.7 $\times$ 57 & ~6.9 $\times$ 57 & 12.2 & 159.2 \\
DLL                & 0.064 & 16.0 & 1.7 $\times$ 13 & ~7.1 $\times$ 13 & 13.5 & ~48.5 \\
BST                & 0.069 & 11.0 & 1.7 $\times$ 18 & ~7.4 $\times$ 18 & 12.2 & ~54.8 \\
Interpreter        & 0.090 & 25.0 & 1.7 $\times$ 45 & 20.8 $\times$ 45 & 41.9 & 222.9 \\
\bottomrule
\end{tabular}
\caption{Verification time for different phases (in seconds). 
Reduction: time used for parsing C and generating straightline paths in Coq. 
Common: time used for compiling common definitions and theorems. 
Avg. compile: the average time used for compiling the straightline Hoare triple definitions of all generated paths $\times$ the number of paths.
Avg. verify: the average time used for verifying each path $\times$ the number of paths.
Max. verify: the maximum time used for verifying each path.
VST verify: the time used for verifying the same programs in VST.
}
\label{tbl:eval_time}
\end{table}

The verification time is shown in Table~\ref{tbl:eval_time}.
Since the splitting process in VST-A is a computationally 
    proven sound function, it is not surprising that the reduction
    time is short.
For different verification tasks, users need to provide different
    separation logic predicates and lemmas to specify and prove
    the program. 
This consitutes a majority of the compilation time, which is also
    needed by a VST proof.

A significant difference between VST-A and VST is that in VST-A 
    each straightline Hoare triple's proof can 
    be developed as a separate lemma, 
    residing in distinct files, 
    facilitating parallel checking and compilation.
Therefore, we present the average time for these phases and the number of paths.
In VST, the correctness theorem of a function needs to be proved 
    as a whole, which is more difficult to parallelize.

In addition to parallelized compilation, the separation of 
    straightline Hoare triple proofs also makes it easier 
    to maintain the correctness of the proofs when the program 
    is modified. Only those paths that are affected
    by the modification need to be re-verified.
We conducted several experiments and the results are shown 
    in Table~\ref{tbl:eval_reverify}.

\begin{table}[hbt]
    \begin{tabular}{llclc}
    \toprule
    Program    & Function        & \multicolumn{1}{l}{Paths} & Changes                   & \multicolumn{1}{l}{Changed paths} \\ 
    \midrule
    Singly linked list & append        & 5    & modify the pre-condition  & 1    \\
    Singly linked list & rev\_append & 4     & change the loop invariant & 3     \\ 
    Singly linked list & reverse      &  4 $\rightarrow$ 3  & remove an assertion in the loop   & 2 $\rightarrow$ 1 \\ 
    Binary search tree & lookup & 5 & modify the post-condition & 2 \\
    Interpreter & eval & 21 & change the code in one branch & 1 \\
    \bottomrule
    \end{tabular}
    \caption{Case study of the number of paths that need to be re-verified when the program is modified.
    }
    \label{tbl:eval_reverify}
\end{table}

\paragraph{Statistics of the development of VST-A}

We present the line of codes statistics for our development in VST-A below. Our development in VST-A includes the following parts:

\begin{itemize}
    \item Coq formalization of the restricted fragment of VST program logic, including the logical rules, auxiliary lemmas, and the conjunction rule proof: 6,236 lines
    \item Coq formalization of the VST-A framework, including the split algorithm, and its soundness proof: 6,287 lines
    \item Modification to the CompCert C parser to parse annotated C programs: 1.87\% of change
    \item Modification to the VST-Floyd lemmas and automation tactics to support forward symbolic execution in the restricted fragment of VST program logic: 6.74\% of change
    \item OCaml development that provides an efficient implementation of the split algorithm: 3,176 lines in addition to original CompCert
\end{itemize}


\section{Related work}
\label{sec:relate}

\subsection{Traditional annotation verifiers}

Many annotation verifiers work
by reducing annotated programs to SMT assertion entailments:
these include Frama-C, Dafny, VeriFast, Viper, Hip/Sleek.
Some of those systems use a specialized intermediate language
for verification, connectable to several SMT back-ends and to several programming-language
front-ends; for example Frama-C uses Why3 \cite{filliatre13:why3} and Dafny uses 
Boogie \cite{barnett06:boogie}.
Modern SMT solvers effectively and efficiently
solve many of the resulting entailments.
These annotation verifiers are, in practice, ``interactive:''
one starts by annotating the program with function specifications and some loop invariants, and the
verifier inevitably points out several places where the proof fails---with
sufficiently good error messages that the user can
adjust the assertions, add new assertions and invariants,
and try again, and again. 
This works well in practice,
and it is what we wanted to emulate.

The disadvantage of those systems is in the poverty of their
assertion languages.  Because SMT (or Why3 or Boogie) accommodates only first-order
or near-first-order logics, the rich specification languages
of VST or of logics built in Iris cannot be used.
In VST one often proves high-level
properties about the behavior of programs, in application-specific
mathematics that would be difficult to fit into SMT.
In Frama-C, Dafny and CN (etc.), some authors work around this by
writing higher-level proofs in Coq and program-logic proofs
in the annotation verifier, and stapling together the two verifications
(in different logics without a common foundation)
\cite{boldo2014}.  That approach could be made more foundational by embedding a semantics of Why3 in Coq \cite{cohen24:why3}, but the assertion language would still be near-first-order.

Another disadvantage of those annotation verifiers is that none has
a machine-checked proof of soundness (e.g., w.r.t. an
operational semantics).  This lack is not
inherent, as VST-A demonstrates.

When proving programs in VST or in an Iris-based logic,
one generally uses automated solvers to prove entailments,
or at least to prove the easy parts and leave residuals for the
user.  These solvers may be programmed in Coq
(using tactics or computational reflection) or may be
external (such as SMT).
VST-A does not choose a specific solver. Users can choose 
their own solver to prove
the split result correct.

BRiCk~\cite{BRiCK}, used by BedRock Systems Inc. to verify
its microkernel/hyperviser, is a program logic for C++, built in Iris, on principles inspired by VST.  We
expect that our VST-A method would work well in such a C++ program logic.

Tools like F*~\cite{Fstar}, 
LiquidHaskell~\cite{LiquidHaskell, LiquidHaskell-PLE},
and ATS~\cite{ATS} have managed to combine higher-order programming with 
theorem proving in a dependent type system so that rich higher-order properties
can be automatically verified  in the style of the program's annotations.
However, they require users to either write programs in a new
domain-specific language or construct the proof as a term in the program.
By contrast, VST-A works on the standard (and practical)
C programming language
while also enabling reasoning with higher-order properties.




We also note the work of sledgehammer~\cite{sledgehammer} and
auto2~\cite{auto2prover} for proof automation in
Isabelle. Sledgehammer relies on SMT solvers,
while auto2 builds compositional proof automation 
using a saturation-based proof automation 
system in which goal-directed proof strategies can be encoded.
Users of auto2 can easily extend auto2 with their own 
domain-specific proof strategy. 
\citet{auto2} built an auto2 instance that supports separation logic 
reasoning for verifying sequential programs. 
Although auto2 supports flexible saturation-based proof 
strategies, this specific instance of sequential program 
verification is mainly goal-directed.
VST-A is not goal-directed, and is open to any solver 
or proof style when verifying entailments,
regardless of whether it is, 
based on an interactive proof
a tactic-based solver, 
or a model checking based one.


\subsection{Interactive prover-based program verification}
There is no deep reason why annotation verifiers should lack soundness proofs (e.g., Frama-C, Dafny, Verifast) and tactic-based verifiers should have machine-checked soundness proofs (e.g., VST, Iris).  We guess that the reason is: such soundness proofs are naturally higher-order, more easily accommodated in the kinds of higher-order logics implemented in proof assistants such as Coq and Isabelle, so it is natural that designers of VST and Iris
\emph{also} have their \emph{users} operate in the same proof assistants.

Soundness proofs are important for real-world programming languages, which have many subtle features in their semantics and compilers.
Users want
what they prove about a program to be consistent 
with the compiled machine code semantics,
so the foundational soundness of
VST-A is a real benefit.
In this section, we compare VST-A with other foundational tools.

VST and various Iris-based verifiers have invested significantly in increasing proof
automation so that users can verify their programs conveniently.
However, they all require their users to complete 
correctness proofs for the entire program in an interactive theorem prover,
 which is not easy for an ordinary software engineer to learn.


There are also works that build annotation verification into
interactive theorem provers and have achieved foundational soundness.
RefinedC~\cite{refinedC} is an automated and foundational
annotation verifier for C programs, that defines a restricted fragment of
the Iris logic, Lithium, so that proof searches can be guided by 
translating the assertion annotations into a Lithium program.
DiaFrame~\cite{DiaFrame} is also an automated and foundational tool. 
It employs a similar structural approach to RefinedC, but is more 
targeted at proving fine-grained concurrent programs.
These tools use tactic-based proof strategy design and achieve some reasonable automation.
In other words, a Hoare triple will be reduced to smaller proof goals (and even directly solved) by automatically applying a series of proof tactics, which use proved-sound logic rules or verified single-step symbolic execution.
In comparison, VST-A is based on one computational proved-sound reduction function.
Thus, developers of VST-A do not need to \emph{decompose} this reduction step into multiple proof tactics, which in the end allows users to describe more flexible proofs using annotated C programs.
Here is an example of how reduction is decomposed into tactics. Given an annotated program of form (here, we use a pair of braces to emphasize that sequential composition is right associative in CompCert Clight and in our ClightA syntax):
\begin{lstlisting}[language=C, mathescape=true, xleftmargin=2em,xrightmargin=1em ]
/*@ $P$ */ if ($b$)  $c_1$  else $c_2$; { $c_3$; /*@ $Q$ */ $c_4$ } /*@ $R$ */
\end{lstlisting}
VST-A will generate 3 straightline Hoare triples:
$$\{P\} \textbf{assume} \ b; c_1; c_3 \{Q\} \quad \quad \{P\} \textbf{assume} \ ! b; c_2; c_3 \{Q\} \quad \text{and} \quad \{Q\} c_4 \{R\}.$$
In order to achieve this in RefinedC's or DiaFrame's tactic-based proof automation system, the system needs to apply the \rulename{Seq-Assoc} rule (see \S\ref{sec:hoarelogic}) first, turning the proof goal into:
\begin{lstlisting}[language=C, mathescape=true, xleftmargin=2em,xrightmargin=1em ]
$\{P\}$ { if ($b$)  $c_1$  else $c_2$; $c_3$ } ;  $c_4$ $\{R\}$
\end{lstlisting}
and then apply the sequence rule with middle condition $Q$.
After that, one more proof rule\footnote{In most Iris-based verifiers, this last step is not needed since Iris's symbolic execution can do that implicitly.} is needed for turning \lstinline[language=C, mathescape=true]{if ($b$)  $c_1$  else $c_2$; $c_3$} into \lstinline[language=C, mathescape=true]{ if ($b$) {  $c_1$; $c_3$ $\}$ else { $c_2$; $c_3$ $\}$ } so that the \rulename{Semax-If} rule can be used to complete the reduction.
However, such tactic-based decomposition is not always easy to design, and it can even be impossible.
Especially, it is not obvious how to design tactic-based proof automation for handling nontraditional loop invariants supported by VST-A.
That is, our split function (in effect) performs some nontrivial static analysis and is verified by a nontrivial soundness proof.
Besides supporting more flexible proofs, the core split function in VST-A is computation-based
so that VST-A can first complete the reduction step very efficiently, and users can then prove straightline
   Hoare triples manually or using their own domain-specific proof automations.
Also, this design of VST-A can better support incremental development.








\subsection{Conjunction rule and preciseness}

The conjunction rule is naturally sound in traditional Hoare logic
~\cite{Floyd1993}. However, for concurrent separation logic with locks,
a counterexample that leads to unsoundness~\cite{OHearn2004} can be found.
As a workaround, ~\citet{DeVilhena2020} proved a restricted version 
called the candidate rule, which requires postconditions to be pure 
(independent of resources, especially ghost resources) to solve their
verification problem. However, this rule cannot be applied in
our setting. In VST-A, we do not propose alternative rules but prove the
conjunction rule on top of a VST logic without ghost updates.
As for supporting concurrency, we proposed several 
possible directions in \S\ref{sec:conj_remark}.

We are not aware of any similar notions of precise function specifications
in the literature
as we have defined in this paper.
Traditionally, preciseness restrictions
are placed on assertion predicates. For example, in concurrent separation
logic, the resource invariant should be a precise predicate 
to make the conjunction rule sound~\cite{Gotsman2011, Vafeiadis2011}.
Our aim is to define a notion of preciseness for specifications,
so that the conjunction rule is derivable from the
existing logical rules.
Compared with traditional preciseness, we showed in 
\S\ref{sec:conj_call} that our notion of preciseness is more expressive,
as it accounts for a pair of pre-/post-conditions for an operation
and allows the specification to be quantified by logical variables.





\section{Conclusion}
\label{sec:concl}

We have presented VST-A, an annotation verifier 
    that is foundationally verified.
VST-A targets a widely used real-world language, C, and supports 
higher-order assertions
in the very rich specification language of VST
that includes the full expressive power of Coq.
VST-A splits the verification of a large program into 
    verifications of straightline control flow paths separated by assertions.
The soundness of this approach requires the conjunction rule
    to be derivable in the program logic. 
We have identified a novel notion of precise specifications in 
    the proof of the conjunction rule.
Currently, VST-A only supports sequential C program verification,
    but we have proposed ways to extend VST-A to support concurrency in the future.
Our formal annotation language and other major designs are not 
    C-specific, nor are they separation-logic-specific, nor VST-specific. 
A similar development can be used to design other 
    Hoare-style annotation verifiers for imperative languages.

Comparing to existing foundational program verification tools built in interactive theorem provers, VST-A has the following advantages:
\begin{itemize}
    \item Annotation-based proof is a more readable way to explain why a program is correct.
    \item Our annotation-based proof language ClightA is expressive enough to describe nonstructural proofs, which cannot be supported systematically using goal-directed tactic-based proof automation;
    \item VST-A is easier to use --- users only need to write assertions in annotations, and use forward symbolic execution to prove straightline Hoare triples. In comparison, users of other tools like VST and Iris need to use different tactics to handle different program structures like if-conditions, recursions and different loops.
    \item VST-A reduces proof recompilation time.
    When a verified program is 
    updated slightly, its correctness proof also needs corresponding updates. 
    In existing interactive verifiers, users must rerun all tactical 
        proof scripts, even though only a small portion needs to be updated. 
    Now, only the part of the program that has been changed and the corresponding proof 
        need to be recompiled, since other parts in $\splitfun$'s result are unchanged.
\end{itemize}

We aim to enhance the verification process of VST-A further. 
Future work includes the introduction of domain-specific heuristics 
    for automatically manipulating separation logic predicates during symbolic execution 
    and proving separation logic entailments on straightline Hoare triples. 
We also plan to allow user to specify partial assertions, so that 
    users do not need to write assertions for the entire state of the function 
    throughout the whole program.


\begin{acks}                            
This material is based upon work supported (in part) by NSF China 61902240, 
and the Defense Advanced Research Projects Agency (DARPA) under Contract 
No. HR001120C0160.
\end{acks}

\bibliography{base}


\newpage

\appendix

\ifhasappendix

\section{Verifiable C's compositional rules} \label{sec:veric-comp-rule}

\begin{figure*}[h]
\centering
\begin{mathpar}
    \inferrule*[left=Semax-Conseq]{
        P_1 \entails P_2 \and
        R_2 \entails R_1 \and
        R'_2 \entails R'_1 \and
        \hoareabbr{P_2}{c}{R_2}{R'_2}
    } {
        \hoareabbr{P_1}{c}{R_1}{R'_1}
    }
    \and
    \inferrule*[left=Semax-Skip]{\,} {
         \hoareabbr{P}{\cskip}{P}{Q}
    }
    \and
    \inferrule*[left=Semax-Break]{\,} {
        \hoare{\brkcnd{Q}}{\cbreak}{Q}{\brkcnd{Q}}{\concnd{Q}}{\retcnd{Q}}
    }
    \and
    \inferrule*[left=Semax-Continue]{\,}{
        \hoare{\concnd{Q}}{\ccontinue}{Q}{\brkcnd{Q}}{\concnd{Q}}{\retcnd{Q}}
    }
    \and
    \inferrule*[left=Semax-Return]{\,}{
        \hoare{\retcnd{Q}}{\creturn}{Q}{\brkcnd{Q}}{\concnd{Q}}{\retcnd{Q}}
    }
    \and
    \inferrule*[left=Semax-Seq]{
        \hoareabbr{P}{c_1}{R}{Q'} \and
        \hoareabbr{R}{c_2}{Q}{Q'}
    } {
        \hoareabbr{P}{c_1;c_2}{Q}{Q'}
    }
    \and
    \inferrule*[left=Semax-If]{
        \hoareabbr{P\land b\neq 0}{c_1}{Q}{Q'} \and
        \hoareabbr{P\land b=0}{c_2}{Q}{Q'}
    } {
        \hoareabbr{P}{\cifthenelse{b}{c_1}{c_2}}{Q}{Q'}
    }
    \and
    \inferrule*[left=Semax-Loop]{
        \hoare{I}{c}{\coninv{I}}{Q}{\coninv{I}}{\retcnd{Q}} \and
        \hoare{\coninv{I}}{\cincr{c}}{I}{Q}{\bot}{\retcnd{Q}}
    } {
        \hoare{I}{\cloop{c}{\cincr{c}}}{Q}{\brkcnd{Q}}{\concnd{Q}}{\retcnd{Q}}
    }
\end{mathpar}
\caption{Compositional rules of C Hoare logic}
\end{figure*}

\section{Computing split result of sequential composition, if statements and loops} \label{sec:split_def}

If \newline
$$
    \splitfun{(C_1)} = \left\{
        \begin{array}{ll}
            \pnormalatom,
            \pbreakatom,
            \pcontinueatom,
            \preturnatom, \\
            \pnormalpost,
            \pbreakpost,
            \pcontinuepost,
            \preturnpost, \\
            \ppre,
            \ppath
        \end{array}
    \right\}
$$
$$
    \splitfun{(C_2)} = \left\{
        \begin{array}{ll}
            \qnormalatom,
            \qbreakatom,
            \qcontinueatom,
            \qreturnatom, \\
            \qnormalpost,
            \qbreakpost,
            \qcontinuepost,
            \qreturnpost, \\
            \qpre,
            \qpath
        \end{array}
    \right\},
$$
then
    $$\begin{array}{l}
    \begin{array}{l}
        \splitfun{(C_1 ; C_2)} = 
        \\
        \quad \left\{
            \begin{array}{l}
                \pnormalatom \conn \qnormalatom, \\
                \pbreakatom \append \pnormalatom \conn \qbreakatom, \\
                \pcontinueatom \append \pnormalatom \conn \qcontinueatom, \\
                \preturnatom \append \pnormalatom \conn \qreturnatom, \\
                \pnormalpost \append \pnormalpost \conn \qnormalatom, \\
                \pbreakpost \append \qbreakpost \append \pnormalpost \conn \qbreakatom, \\
                \pcontinuepost \append \qcontinuepost \append \pnormalpost \conn \qcontinueatom, \\
                \preturnpost \append \qreturnpost \append \pnormalpost \conn \qreturnatom, \\
                \ppre \append \pnormalatom \conn \qpre, \\
                \ppath \append \qpath \append \pnormalpost \conn \qpre
            \end{array}
        \right\}
    \end{array}
    \quad
    \begin{array}{l}
        \splitfun{(\cifthenelse{e}{C_1}{C_2})} =\\
        \quad \left\{
            \begin{array}{l}
                \singleton{ [\assumee] } \conn \pnormalatom \append \singleton{ [\assumenote] } \conn \qnormalatom \\
                \singleton{ [\assumee] } \conn \pbreakatom \append \singleton{ [\assumenote] } \conn \qbreakatom \\
                \singleton{ [\assumee] } \conn \pcontinueatom \append \singleton{ [\assumenote] } \conn \qcontinueatom \\
                \singleton{ [\assumee] } \conn \preturnatom \append \singleton{ [\assumenote] } \conn \qreturnatom \\
                \pnormalpost \append \qnormalpost \\
                \pbreakpost \append \qbreakpost \\
                \pcontinuepost \append \qcontinuepost \\
                \preturnpost \append \qreturnpost \\
                \singleton{ [\assumee] } \conn \ppre \append \singleton{ [\assumenote] } \conn \qpre, \\
                \ppath \append \qpath
            \end{array}
        \right\}
    \end{array}
    \\
    \\
    \begin{array}{ll}
        \splitfun{(\cloop{C_1}{C_2})} = & \textsf{if } (\pnormalatom \append \pcontinueatom) \conn \qnormalatom = \nil \textsf{ then } \\
        & \left\{
            \begin{array}{l}
                \pbreakatom
                    \append (\pnormalatom \append \pcontinueatom) \conn \qbreakatom, \\
                \nil, \\
                \nil, \\
                \preturnatom 
                    \append (\pnormalatom \append \pcontinueatom) \conn \qreturnatom, \\
                \pbreakpost \append \qbreakpost
                    \append (\pnormalpost \append \pcontinuepost \append \qnormalpost \conn (\pnormalatom \append \pcontinueatom)) \conn \qbreakatom \\ \qquad
                    \append (\qnormalpost \append (\pnormalpost \append \pcontinuepost) \conn \qnormalatom ) \conn \pbreakatom, \\
                \nil, \\
                \nil, \\
                \preturnpost \append \qreturnpost
                    \append (\pnormalpost \append \pcontinuepost \append \qnormalpost \conn (\pnormalatom \append \pcontinueatom)) \conn \qreturnatom \\ \qquad
                    \append (\qnormalpost \append (\pnormalpost \append \pcontinuepost) \conn \qnormalatom )\conn \preturnatom, \\
                \ppre 
                    \append (\pnormalatom \append \pcontinueatom) \conn (\qpre \append \qcontinueatom \conn \singleton{ []\postassert{\bot}}), \\
                \ppath 
                    \append \qpath
                    \append \qnormalpost \conn \ppre \append \qcontinuepost  \conn \singleton{ []\postassert{\bot}} \\ 
                    \quad \append (\pnormalpost \append \pcontinuepost \append \qnormalpost \conn (\pnormalatom \append \pcontinueatom))  \\
                    \qquad\quad \conn (\qpre \append \qnormalatom \conn \ppre \append \qcontinueatom \conn \singleton{ []\postassert{\bot}} )
                
            \end{array}
        \right\}
    \end{array}

\end{array}$$

\section{Coq dependent type for defining split results}\label{sec:exgiven_dt}

A technical issue arises when implementing the split function for the $\ExGivensymbol$ structure in Coq.
To allow unrestricted logical variables in assertions and proofs,
we encoded the quantification of the logical variables as shallow-embedded 
function types in Coq, i.e. the Coq type of ExGiven can be written as:
\begin{mdframed}[
    backgroundcolor=gray!10,
    leftmargin=0.5cm,skipabove=0.1cm,hidealllines=true,%
    innerleftmargin=0.2cm,innerrightmargin=0.2cm,
    innertopmargin=-0.0cm,innerbottommargin=-0.0cm]
\begin{lstlisting}[language=Caml, 
    basicstyle=\sffamily,
    keepspaces=true, 
    aboveskip=0.25em,
    belowskip=0.25em,
    columns=flexible,
    keepspaces=true,frame = {H}]
  ExGiven'              : forall A,(A -> assert)->(A -> ClightA_stmt)->ClightA_stmt
\end{lstlisting}
\end{mdframed}
If we want to split the ClightA statement
$\ExGiven{x}{A}{P(x)}{C'}$, where $x$ is a logical variable with Coq type $A$,
we can at most compute the Coq function that accepts a variable $x$ of type $A$
and returns the entire split result of $C'$ instantiated with $x$, which is indicated
by the following Coq term:
\begin{mdframed}[
    backgroundcolor=gray!10,
    leftmargin=0.5cm,skipabove=0.1cm,hidealllines=true,%
    innerleftmargin=0.2cm,innerrightmargin=0.2cm,
    innertopmargin=-0.0cm,innerbottommargin=-0.0cm]
\begin{lstlisting}[language=Caml, 
    basicstyle=\sffamily,
    keepspaces=true, 
    aboveskip=0.25em,
    belowskip=0.25em,
    columns=flexible,
    keepspaces=true,frame = {H}]
  fun x => split(C' x)  : A -> split_result.
\end{lstlisting}
\end{mdframed}
With shallow-embedded logical variables,
    the above Coq type cannot
    guarantee that the split results are in the same shape
    when we instantiate the term with different $x$'s,
but in fact, different instantiations of $x$ only change
    the assertion part and do not
    affect the paths in the split result.
We use Coq dependent types to enforce this restriction.
Correspondingly, the ClightA program must also be restricted
    by dependent types to state that the parameter of
    the $\textsf{ExGiven}$ constructor takes effect only on
    the assertion but not the program.
Specifically, we define the ClightA syntax's Coq types 
    and the intermediate split results as follows:
\begin{mdframed}[
    backgroundcolor=gray!10,
    leftmargin=0.5cm,skipabove=0.1cm,hidealllines=true,%
    innerleftmargin=0.2cm,innerrightmargin=0.2cm,
    innertopmargin=-0.0cm,innerbottommargin=-0.0cm]
\begin{lstlisting}[language=Caml, 
    basicstyle=\sffamily,
    keepspaces=true, 
    aboveskip=0.25em,
    belowskip=0.25em,
    columns=flexible,
    keepspaces=true,frame = {H}]
  simple_ClightA_stmt   : Type
  ClightA_stmt          : simple_ClightA_stmt -> Type
  ExGiven               : forall (S: simple_ClightA_stmt) (A: Type),
                            (A -> assert) -> (A -> ClightA_stmt S) -> ClightA_stmt S
  simple_split_result   : Type
  split_result          : simple_split_result -> Type
\end{lstlisting}
\end{mdframed}
The simpler ClightA syntax and intermediate split result 
erase both the quantifiers and the contents of the assertions, and they only preserve 
a place-holder for every assertion, as presented in \figref{fig:SimpleAST}.
Therefore, shallow-embedded function types no longer appear in
simpler syntax trees and simpler split results.
Now, the complete split result of inner statement $C'$ 
has the dependent type of a simpler result, which is not abstracted by the logical variable's type $A$, 
so we are able to perform pattern-matching on that simpler result,
and extract individual paths from the abstracted split result.
The implemented \textsf{split} function has the following Coq type:
\begin{mdframed}[
    backgroundcolor=gray!10,
    leftmargin=0.5cm,skipabove=0.1cm,hidealllines=true,%
    innerleftmargin=0.2cm,innerrightmargin=0.2cm,
    innertopmargin=-0.0cm,innerbottommargin=-0.0cm]
\begin{lstlisting}[language=Caml, 
    basicstyle=\sffamily,
    keepspaces=true, 
    aboveskip=0.25em,
    belowskip=0.25em,
    columns=flexible,
    keepspaces=true,frame = {H}]
  simple_split          : simple_ClightA_stmt -> simple_split_result
  split                 : forall (S: simple_ClightA_stmt), 
                            ClightA_stmt S -> split_result(simple_split S)
\end{lstlisting}
\end{mdframed}

\begin{figure}[h]
    $$
    \begin{array}{c}
    \begin{array}{rrl}
        \text{Simple ClightA}:  &C_s :=& \ldots \text{ (same as $\ldots$ in $C$) } \mid \cassert{} \\
        \text{ClightA}: &C :=& \ldots   \mid \cassert{P}  \ \mid \  \ExGiven{x}{A}{P(x)}{C}
    \end{array} \\
    \\
    \begin{array}{rrl}
        \text{Simple head path}:     &p_{s\prenotation} :=&  \vec{c_b} \postassert{} \\
        \text{Head path}:    &p_{\prenotation} :=&  \vec{c_b} \postassert{P} \\
        \text{Simple tail path}:    &p_{s\postnotation} :=&  \preassert{} \vec{c_b} \\
        \text{Tail path}:   &p_{\postnotation} :=&  \preassert{P} \vec{c_b} \\
        \text{Simple full path}:            &p_{s\pathnotation} :=&  \preassert{} \vec{c_b}  \postassert{} \\
        \text{Full path}:           &p_{\pathnotation} :=&  \preassert{P_1} \vec{c_b}  \postassert{P_2} \
                                                            \mid \  \forall (x:A).\ {p_{\pathnotation}} \\
    \end{array}
    \end{array}
    $$
        \caption{Dependent type definition of ClightA programs and split results. \\ 
        {\footnotesize In ClightA, $\cassert{P}$ is of type $\textsf{assert}$, and $\ExGiven{x}{A}{P(x)}{C}$ is of type
        ``$\textsf{assert};C_s$'' when $C$ is of type $C_s$. \\
        In the split results, $p$ is a path of dependent type $p_s$ if $p$ and $p_s$ have the same shape and basic statements
        }}
        \label{fig:SimpleAST}
    \end{figure}

\section{Soundness proof for splitting if/loop} \label{sec:if_loop_sound}

\paragraph{Proof of Proposition \ref{prop:semax_seq_inv_split_group}}

Assuming conjunction rule, the following propositions can be proved:

\begin{proposition}[Conjunction rule on paths]
    \label{prop:conj_path}
    \,
    \begin{enumerate}
        \item If $\semaxpost(Q_1, q_{\postnotation})$ and $\semaxpost(Q_2, q_{\postnotation})$, 
            then $\semaxpost(Q_1 \land Q_2, q_{\postnotation})$
        \item If $\semaxatom(P, Q_1, q_{\atomnotation})$ and $\semaxatom(P, Q_2, q_{\atomnotation})$,
             then $\semaxatom(P, Q_1 \land Q_2, q_{\atomnotation})$
        \item If $\semaxposta(Q_1, \qpost)$ and $\semaxposta(Q_2, \qpost)$,
            then $\semaxposta(Q_1 \land Q_2, \qpost)$
        \item If $\semaxatoma(P, Q_1, \patom)$ and $\semaxatoma(P, Q_2, \patom)$,
             then $\semaxatoma(P, Q_1 \land Q_2, \patom)$
    \end{enumerate}
\end{proposition}

Proposition \ref{prop:semax_seq_inv_split_group} can be proved by induction fisrt on the
size of set on the left argument of the $\conn$ operator then on the size of set on the right argument.
The induction on the left argument is straightforward, since the intermediate assertions
we obtain from the inversion lemmas are the same if we fix the right argument.
Consider the case for the second induction when proving 
Proposition \ref{prop:semax_seq_inv_split_group} (\ref{enum:semaxpost_inv_group}).
Assume $\qatom = q_{\atomnotation} :: \boldsymbol{q'}_{\atomnotation}$. Applying
Proposition \ref{prop:semax_seq_inv_split} (\ref{enum:semaxpost_inv}) to $q_{\atomnotation}$
we get $$\semaxposta(\EXabbr{R_1}{\semaxpre (R_1, q_{\atomnotation})}, \ppost)$$. The induction
hypothesis gives $$\semaxposta(\EXabbr{R_2}{\semaxprea(R_2, \boldsymbol{q'}_{\atomnotation})}, \ppost)$$.
Let $R$ be $R_1 \land R_2$, we can derive that
$$
\begin{array}{rl}
    & \EXabbr{R_1}{\semaxpre (R_1, q_{\atomnotation})}
    \land \EXabbr{R_2}{\semaxprea(R_2, \boldsymbol{q'}_{\atomnotation})} \\
    \entails & \EXabbr{R}{\semaxprea(R, \qatom)}
\end{array}
$$

Applying Proposition \ref{prop:conj_path}, we can get the LHS of the above derivation
and finish the proof.

\paragraph{Soundness of splitting loop}
Similar to sequencing case, for soundness of $\splitfun{(\cloop{c_1}{c_2})}$, we construct the loop 
invariant $I$ and the invariant for the incremental step $\coninv{I}$ required
by $\textsc{Semax-Loop}$ as follows:

$$
I = \left\{\begin{array}{l}
    \exists R. \ R \land \semaxprea(R, \ppre) \\
        \land \ \semaxprea(R, (\pnormalatom \append \pcontinueatom) \conn \qpre) \\
        \land \ \semaxatoma(R, \bot, (\pnormalatom \append \pcontinueatom) \conn \qcontinueatom  ) \\
        \land \ \semaxatoma(R, Q, \pbreakatom) \\
        \land \ \semaxatoma(R, Q, (\pnormalatom \append \pcontinueatom) \conn  \qbreakatom) \\
        \land \ \semaxatoma(R, \retcnd{Q}, \preturnatom) \\
        \land \ \semaxatoma(R, \retcnd{Q}, (\pnormalatom \append \pcontinueatom) \conn  \qreturnatom)
\end{array}\right\}
\,
\coninv{I} = \left\{\begin{array}{l}
    \exists R. \ R \land \semaxprea(R, \qpre) \\
        \land \ \semaxprea(R, \qnormalatom \conn \ppre) \\
        \land \ \semaxatoma(R, \bot, \qcontinueatom) \\
        \land \ \semaxatoma(R, Q, \qbreakatom) \\
        \land \ \semaxatoma(R, Q, \qnormalatom \conn \pbreakatom) \\
        \land \ \semaxatoma(R, \retcnd{Q}, \qreturnatom) \\
        \land \ \semaxatoma(R, \retcnd{Q}, \qnormalatom \conn \preturnatom)
\end{array}\right\}
$$

By inversion from the premise that 
$$\semaxsplit(P, Q, \brkcnd{Q}, \concnd{Q}, \retcnd{Q}, \splitfun({\cloop{c_1}{c_2}}))$$,
we can derive the two Hoare triples required by $\textsc{Semax-Loop}$ and
prove $\splitfun{(\cloop{c_1}{c_2})}$ sound.

\paragraph{Soundness of if-branching}
The key for this case is to show 
that it is safe to translate the \textsf{assume} statement into the pre-condition.
Recall the encoding we have defined for $\textsf{assume } e$:

$$ \textsf{assume} \ e \ \triangleq \textbf{if} \ (e) \ \text{break} \ \textbf{else} \ \text{skip};$$.

From lemma \ref{thm:assume_fact}, we can derive the following propositions (the dual case for $\textsf{assume} \ \neg e$ omitted):

\begin{proposition}[Conditional Expression lemmas]
    \label{prop:expr_check_lemmas}
    \,
    \begin{enumerate}
        \item \label{enum:expr_check} If $\semaxprea(P, \singleton{ [\textsf{assume} \ e] }  \conn \ppre)$, then
              $\semaxprea(P \land e \neq 0 , \ppre)$
        \item If $\semaxatoma(P, Q, \singleton{ [ \textsf{assume} \ e] }  \conn \patom)$, then
             $\semaxatoma(P\land e \neq 0 , Q, \patom)$
    \end{enumerate}
\end{proposition}

After transformation using Lemma \ref{lemma:inversion_if} and 
Proposition \ref{prop:expr_check_lemmas} on the premise of the soundness theorem, 
the induction hypothesis can be directly applied 
to complete the proof.

\section{Preciseness proof for memory store}

We use $\join$ to denote the join relations in separation algebra.
The semantic model (known as resource map) of VST, and the permissions
are two instances of the separation algebra. Due to the existence
of ghost states, the join relation on resource map does not enjoy
some properties of the join relation on permissions, such as
the cross split property and the cancellative property.

\begin{proposition}[Cross Split Property] A join relation for
    permissions has the cross split property if
    $\pi = \pi_a \join \pi_b \land 
    \pi = \pi_c \join \pi_d \implies \exists \pi_{ac} \ \pi_{ad} \ 
    \pi_{bc} \ \pi_{bd}, \pi_{ac} \join \pi_{ad} = \pi_a \land
    \pi_{bc} \join \pi_{bd} = \pi_b \land
    \pi_{ac} \join \pi_{bc} = \pi_c \land
    \pi_{ad} \join \pi_{bd} = \pi_d$.
\end{proposition}

\begin{proposition}[Cancellative Property]
    A join relation for
    permissions has the cancellative property if
    $\pi = \pi_a \join \pi_b \land 
    \pi = \pi_{a'} \join \pi_b \implies \pi_a = \pi_{a'}$.
\end{proposition}

Therefore, proofs for memory loading/writing will look into
the location of the value being loaded/stored on the resource map,
so that we can make use of properties of the join relation on
permissions. We use $\resat{r}{l}$ to refer to the resource
on location $l$ of the model $r$.

\begin{lemma} 
    \label{thm:load_conj}
    If $\pi_1$ and $\pi_2$ are readable,
    then $$
    \begin{array}{rl}
        &(p \mapsto_{\pi_1} v_1 \sepcon \TT) \land 
        (p \mapsto_{\pi_2} v_2 \sepcon \TT) \\
        \entails &
         v_1 = v_2 
        \land  p \mapsto_{\pi_1 \cup \pi_2} v_1 \sepcon \TT
    \end{array}
    $$
\end{lemma}
\begin{proof}

    Given a model $r$ that satisfy the LHS of the theorem, by the
    semantics of $\sepcon$, there are two ways to disjointly split
    this model, say $r = r_1 \join r'_1$ and $r = r_2 \join r'_2$, where
    $r_1 \models p \mapsto_{\pi_1} v_1$ and $r_2 \models p \mapsto_{\pi_2} v_2$.

    To show $v_1 = v_2$, it suffices to show that the resources
    on location $l$ referenced by $p$ in $r_1$ and $r_2$ are equal.
    This is done by inversion on the join relation since both
    $r_1$ and $r_2$ are part of the same model $r$.

    As for the second part, we can pointwisely
    define the two models $r_0$, $r'_0$ that constitute 
    $p \mapsto_{\pi_1 \cup \pi_2} v_1 \sepcon \TT$. 
    
    For locations $l$ referenced by $p$, let the permission of $\resat{r}{l}$
    be $\pi$.  From the cross split property of permissions, we have
    $(\pi_1 \cup \pi_2) \join (\pi \cap (\neg (\pi_1 \cup \pi_2))) = \pi$.
    We can define $\resat{r_0}{l}$ and $\resat{r'_0}{l}$ to
    hold the two sub permissions as above respectively. For other locations,
    we can simply take $\resat{r_0}{l}$ and $\resat{r'_0}{l}$
    to be $\resat{r_1}{l}$ and $\resat{r'_1}{l}$. It follows that
    $r_0 \join r'_0 = r$ and $r_0 \models p \mapsto_{\pi_1 \cup \pi_2} v_1$.    
    
\end{proof}

\begin{theorem}[Preciseness of store]
    \label{thm:write_conj_app}
    \begin{displaymath}
        \begin{array}{rl}
            & (\exists \ v_1 \ \pi_1. \ \pi_1 \text{ is writable} \ \land \  p \mapsto_{\pi_1} v_1 \sepcon (p \mapsto_{\pi_1} v' \wand P_1)) \ \land \\
            & (\exists \ v_2 \ \pi_2. \ \pi_2 \text{ is writable} \ \land  \  p \mapsto_{\pi_2} v_2 \sepcon( p \mapsto_{\pi_2} v' \wand P_2)) \\
            \entails & \exists \ v \ \pi. \  \pi \text{ is writable} \ \land  \   p \mapsto_\pi v \sepcon  (p \mapsto_{\pi} v' \wand P_1 \land P_2)
        \end{array}
    \end{displaymath}
\end{theorem}
\begin{proof}
    For any model $r$ that satisfies the LHS of the derivation,
    there are two ways to disjointly split this model, say $r = r^{\pi_1}_{\mapsto} \join r^{\pi_1}_{\text{rem}}$
    and $r = r^{\pi_2}_{\mapsto} \join r^{\pi_2}_{\text{rem}}$, where
    $r^{\pi_1}_{\mapsto} \models p \mapsto_{\pi_1} v_1$ and 
    $r^{\pi_2}_{\mapsto} \models p \mapsto_{\pi_2} v_2$.

    Applying Lemma \ref{thm:load_conj}, we can show that $v_1 = v_2$
    and that there exists
    a splitting for $r = r^{\pi_1 \cup \pi_2}_{\mapsto} \join r^{\pi_1 \cup \pi_2}_{\text{rem}}$ where
    $r^{\pi_1 \cup \pi_2}_{\mapsto} \models p \mapsto_{\pi_1 \cup \pi_2} v_1$.

    We are left to prove that given any model $r'_{\mapsto}$ that satisfies
    $p \mapsto_{\pi_1 \cup \pi_2} v_1$, $r'_{\mapsto} \join r^{\pi_1 \cup \pi_2}_{\text{rem}}$ satisfies $P_1 \land P_2$. Without loss of generality, we show that 
    $r'_{\mapsto} \join r^{\pi_1 \cup \pi_2}_{\text{rem}} = r' \models P_1$.
    \figref{fig:write_conj_split} plots the layout of the join relations that
    have been introduced so far in the proof.

    To prove a model satisfying $P_1$, we must make use of the fact that
    $r^{\pi_1}_{\text{rem}} \models p \mapsto_{\pi_1} v' \wand P_1$
    The only way to relate $r'$ with $r^{\pi_1}_{\text{rem}}$ is through
    $r_{\text{rem}}^{\pi_1 \cup \pi_2}$, the common sub-model of $r'$
    and $r$. The idea is to ``borrow'' a sub-model from $r'_{\mapsto}$,
    to supplement the model $r^{\pi_1 \cup \pi_2}_{\text{rem}}$,
    so that the supplement can be matched with $r^{\pi_1}_{\text{rem}}$.

    Based on this observation, we construct two models, which are defined as
    $$
    \begin{array}{l}
        \resat{r_{\mapsto}^{\pi'_1}}{l} = \left\{\begin{array}{ll}
            [\resat{r'_{\mapsto}}{l} / \pi_1]  & \text{if $l$ referenced by $p$} \\
            \resat{r'_{\mapsto}}{l} & \text{otherwise}
        \end{array}  \right. \\
        \resat{r_{\mapsto}^{\pi_2/\pi_1}}{l} = \left\{ \begin{array}{ll}
            [\resat{r'_{\mapsto}}{l} / (\pi_2/\pi_1)]  & \text{if $l$ referenced by $p$} \\
           \bot & \text{otherwise}
        \end{array}  \right.
    \end{array}
    $$
    , where $[\resat{r}{l}/\pi]$ indicates a resource on location $l$ that has the
    same value as $\resat{r}{l}$ but with the permissions reassigned as $\pi$.
    The two dotted join relations in \figref{fig:write_conj_split} can be verified.

    \begin{enumerate}
        \item $r'_{\mapsto} = r_{\mapsto}^{\pi'_1} \join r_{\mapsto}^{\pi_2/\pi_1}$ follows
                directly from the fact that $\pi_1 \cup \pi_2 = \pi_1 \join (\pi_2/\pi_1)$.
        \item $r_{\text{rem}}^{\pi_1} = r_{\mapsto}^{\pi_2/\pi_1} \join r_{\text{rem}}^{\pi_1 \cup \pi_2}$
            requires inversion on 
            $r = r_{\text{rem}}^{\pi_1} \join r_{\mapsto}^{\pi_1}$ and
            $r = r_{\text{rem}}^{\pi_1 \cup \pi_2} \join r_{\mapsto}^{\pi_1 \cup \pi_2}$.

            For $l$ referenced by $p$, let the permissions for
            $\resat{r}{l}$, $\resat{r_{\text{rem}}^{\pi_1}}{l}$ and
            $\resat{r_\text{rem}^{\pi_1 \cup \pi_2}}{l}$ be $\pi$, $\pi_3$ and $\pi_4$
            respectively. Then, we have $\pi = \pi_3 \join \pi_1$ and 
            $\pi = \pi_4 \join (\pi_1 \cup \pi_2)$. To prove
            $\resat{r_{\text{rem}}^{\pi_1}}{l} = 
            \resat{r_{\mapsto}^{\pi_2/\pi_1}}{l} \join \resat{r_{\text{rem}}^{\pi_1 \cup \pi_2}}{l}$,
            it suffices to show that $\pi_3 = (\pi_2 / \pi_1) \join \pi_4$. The cancellative
            property is used here. 
            
            For $l$ not referenced by $p$,
            no resources are defined on $r_{\mapsto}^{\pi_1}$ and $r_{\mapsto}^{\pi_1\cup\pi_2}$,
            so $r_{\text{rem}}^{\pi_1}$ and $r_{\text{rem}}^{\pi_1\cup\pi_2}$ are the same.
            Since we define no resources for $r_{\mapsto}^{\pi_2/\pi_1}$, the join relation holds.
    \end{enumerate}

    Based on the two join relations discovered above, we can use
    the associativity property of the join relation to show
    $r' = r_{\text{rem}}^{\pi_1} \join r_{\mapsto}^{\pi'_1}$.
    Then $r' \models P_1$ follows from the premise that
    $r^{\pi_1}_{\text{rem}} \models p \mapsto_{\pi_1} v' \wand P_1$.

\begin{figure}

\begin{tikzcd}[row sep=0.6em, column sep=0.3em]
    &                                                                      & r' \arrow[ldd, no head] \arrow[rrrddd, no head] &  &                                                                                   &                                   & r \arrow[lld, no head] \arrow[rd, no head] \arrow[lddd, no head] \arrow[rddd, no head] &                                \\
    &                                                                      &                                                 &  & r_{\text{rem}}^{\pi_1} \arrow[rdd, no head, dashed] \arrow[lldd, no head, dashed] &                                   &                                                                                        & r_{\mapsto}^{\pi_1}            \\
    & r'_{\mapsto} \arrow[rd, no head, dashed] \arrow[ld, no head, dashed] &                                                 &  &                                                                                   &                                   &                                                                                        &                                \\
r_{\mapsto}^{\pi'_1} &                                                                      & r^{\pi_2/\pi_1}_{\mapsto}                       &  &                                                                                   & r_{\text{rem}}^{\pi_1 \cup \pi_2} &                                                                                        & r_{\mapsto}^{\pi_1 \cup \pi_2}
\end{tikzcd}
\caption{Semantic model layout for the proof of Theorem \ref{thm:write_conj_app}.
(Dotted lines indicate the join relation to be proved)
\label{fig:write_conj_split}}
\end{figure}

\end{proof}




\fi

\end{document}